\documentclass[aps,pre,amsmath,amssymb,longbibliography,lengthcheck,showpacs,superscriptaddress,floatfix]{revtex4-1}
\usepackage{graphicx}
\begin{document}
\title{Beating the amorphous limit in thermal conductivity by superlattices design}
\author{Hideyuki Mizuno}
\email{Hideyuki.Mizuno@dlr.de}
\altaffiliation{Current address: Institut f\"{u}r Materialphysik im Weltraum, Deutsches Zentrum f\"{u}r Luft- und Raumfahrt (DLR), 51170 K\"{o}ln, Germany}
\affiliation{Univ. Grenoble Alpes, LIPHY, F-38000 Grenoble, France}
\affiliation{CNRS, LIPHY, F-38000 Grenoble, France}
\author{Stefano Mossa}
\email{stefano.mossa@cea.fr}
\affiliation{Univ. Grenoble Alpes, INAC-SPRAM, F-38000 Grenoble, France}
\affiliation{CNRS, INAC-SPRAM, F-38000 Grenoble, France}
\affiliation{CEA, INAC-SPRAM, F-38000 Grenoble, France}
\author{Jean-Louis Barrat}
\email{jean-louis.barrat@ujf-grenoble.fr}
\affiliation{Univ. Grenoble Alpes, LIPHY, F-38000 Grenoble, France}
\affiliation{CNRS, LIPHY, F-38000 Grenoble, France}
\affiliation{Institut Laue-Langevin - 6 rue Jules Horowitz, BP 156, 38042 Grenoble, France}
\date{\today}
\begin{abstract}
The value measured in the amorphous structure with the same chemical composition is often considered as a lower bound for the thermal conductivity of any material: the heat carriers are strongly scattered by disorder, and their lifetimes reach the minimum time scale of thermal vibrations. An appropriate design at the nano-scale, however, may allow one to reduce the thermal conductivity even below the amorphous limit. In the present contribution, using molecular-dynamics simulation and the Green-Kubo formulation, we study systematically the thermal conductivity of layered  phononic materials (superlattices), by tuning different parameters that can characterize such structures. We discover that the key to reach a lower-than-amorphous thermal conductivity is to block almost completely the propagation of the heat carriers, the superlattice phonons. We demonstrate that a large mass difference in the two intercalated layers, or weakened interactions across the interface between layers result in materials with very low thermal conductivity, below the values of the corresponding amorphous counterparts.
\end{abstract}
%
%
\maketitle
\section{Introduction}
Materials with low thermal conductivity, $\kappa$, are employed in many modern technologies, such as thermal management in electronic devices or thermoelectric energy conversion~\cite{Venkatasubramanian_2001,Minnich_2009,Maldovan_2013}. In general, low values of $\kappa$ are observed in disordered solids~\cite{Goodson_2007}, including topologically disordered systems (glasses) and crystalline solids with size or mass disorder (disordered alloys)~\cite{Cahill_1988,Cahill_1992,Allen_1993,Mizuno2_2013,Mizuno2_2014}. This behaviour can be rationalized by considering the phenomenological kinetic theory expression~\cite{kettel} $\kappa = (1/3) C v^2 \tau$, which relates average velocity, $v$, and lifetime, $\tau$, (and therefore the mean free path $\ell=v \tau$) of phonons to $\kappa$ ($C$ is the specific heat per unit volume). In good crystals, phonons lifetime is primarily controlled by anharmonic interactions. In contrast, in disordered solids, the disorder (or the elastic heterogeneity~\cite{Mizuno_2013}) reduces $\tau$ (or $\ell$) and, as a result, $\kappa$. 

In early experimental investigations~\cite{Cahill_1988,Cahill_1992}, Cahill {\it et al.} have studied the disordered alloys, e.g., $(\text{KBr})_{1-x}(\text{KCN})_x$, and shown that $\kappa$ can be reduced to the glass value by controlling the relative composition $x$. In our works~\cite{Mizuno2_2013,Mizuno2_2014} we in turn demonstrated that, in size-disordered crystal, $\kappa$  progressively decreases with increasing size mismatch, eventually converging to the corresponding glass value. When this limit is reached, $\tau$ is comparable to the time scale of thermal vibrations ($\ell$ to the particle size), i.e., to the minimum time (length) scale~\cite{Mizuno2_2013,Mizuno2_2014}. Heat propagation can therefore be described as a random walk of vibrational energies~\cite{Cahill_1988,Cahill_1992}, or in terms of non-propagating delocalized modes, the diffusons~\cite{Allen_1993}. For this reason, the value in the glass is generally considered as a lower bound for $\kappa$ of materials with homogeneous chemical composition~\cite{Cahill_1988,Cahill_1992}.

\begin{figure}[b]
\centering
\includegraphics[width=0.48\textwidth]{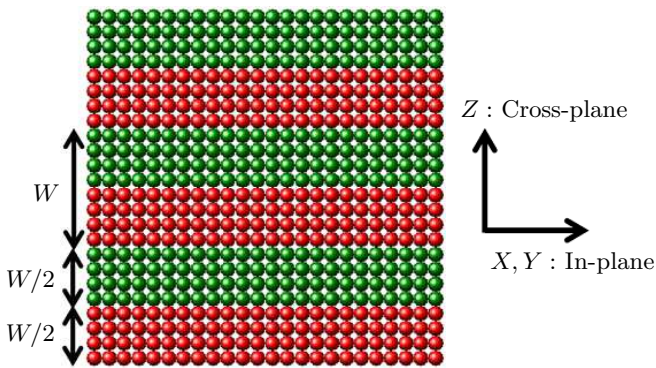}
\caption{
{\bf Schematic illustration of the considered superlattice structures.} The investigated superlattice is composed of two FCC-crystalline layers, $A$ (red) and $B$ (green). The two layers have identical  thickness $w/2$, where $w$ is the replication period. Here, we measure $w$ as the number of monolayers of the crystalline lattice, e.g., $w=8$ in the displayed case. The distance between adjacent monolayers is $a/2$ for theperfect FCC structure we consider, where $a$ is the lattice constant.
} 
\label{schematic}
\end{figure}

A crucial issue~\cite{Goodson_2007} is whether thermal conductivity can be lowered below the glass limit through nanoscale phononic design~\cite{Hopkins_2011,Maldovan_2013}. This possibility would allow to devise (meta-)materials which are excellent thermal insulators while preserving good electronic properties, as needed in many applications~\cite{Venkatasubramanian_2001,Minnich_2009,Maldovan_2013}. The most popular design to reach this goal is that of a lamellar superlattice~\cite{Lee_1997,Volz2_2000,Capinski_1999,Daly_2002,Venkatasubramanian_2000}, often composed of two chemically different intercalated layers, e.g., Si-Ge~\cite{Lee_1997,Volz2_2000} or GaAs-AlAs~\cite{Capinski_1999,Daly_2002} (see also Fig.~\ref{schematic}). In a superlattice, the thermal conductivity tensor is anisotropic, with the cross-plane component, $\kappa_\text{CP}$, usually lower than the in-plane value, $\kappa_\text{IP}$~\cite{Yang_2002,Mavrokefalos_2007}. In recent experiments~ \cite{Costescu_2004,Chiritescu_2007,Pernot_2010}, ultra-low values of $\kappa_\text{CP}$, suggested to be smaller than the amorphous limit, were measured. In particular, Costescu {\em et al.}~\cite{Costescu_2004} demonstrated that the presence of a high-density of interfaces decreases $\kappa_\text{CP}$ of W-$\text{Al}_2\text{O}_3$ nanolaminates, below that of the amorphous $\text{Al}_2\text{O}_3$. An experiment by Chiritescu {\it et al.}~\cite{Chiritescu_2007} achieved ultra-low thermal conductivity in layered $\text{WSe}_2$ crystals, by disordering the crystalline $\text{WSe}_2$ sheets. Finally, Pernot {\it et al.}~\cite{Pernot_2010} also observed very low values of $\kappa_\text{CP}$, below that of amorphous Si, in Ge nanodots multi-layers separated by Si crystals. 
\begin{table*}[t]
\centering
\caption{
{\bf The investigated superlattice structures.}
Details of the three superlattice systems investigated in this work. They are based on the FCC-crystal lattice structure and are composed of: ({\em S1}) two intercalated crystalline layers ($A$ and $B$) formed by sphere particles with different masses $m_A$ and $m_B$; ({\em S2}) ordered crystalline layers intercalated to mass-disordered alloy layers; and ({\em S3}) identical crystalline layers with modified (weakened compared to those intra-layers) interactions across the interfaces. The control parameters are the mass ratio $m_B/m_A$ in {\em S1}, the mass ratio $m_{B2}/m_{B1}$ of the disordered alloy layer in {\em S2}, and the energy scale $\epsilon_{AB}$ of the interactions across the interfaces in {\em S3}. Number density and temperature were fixed to the values $\hat{\rho}=1.015$ (corresponding to a lattice constant $a=1.58$) and $T=10^{-2}$, respectively. The quantities presented in the table are defined in the main text. In the last column we refer to the figure containing the data relative to the indicated system. Additional details about the investigated superlattices and parameters used are given in the {\bf Methods} section.
} 
\label{case}
\vspace*{2.5mm}
\renewcommand{\arraystretch}{1.1}
\begin{tabular}{ccl|ccccccccc|cc}
\hline
\hline
 System & Control Parameter & & & $\kappa_A$ & $\kappa_B$ & $\kappa_{\text{CP}}^{\infty}$ & $\kappa_{\text{IP}}^{\infty}$ & $R$ & $\ell_K$ & $\kappa_{\text{glass}}$ & $\kappa_{\text{disorder}}$ & & Fig. \\
\hline
({\em S1}) Mass difference & $m_B/m_A=$ & $2$ & & $488.6$ & $335.4$ & $397.8$ & $412.0$ &  $0.5$ &  $398$ & $5.7$ & $20.4$ & & Fig.~\ref{kmass}(a) \\
                    &            & $4$ & & $625.9$ & $306.8$ & $411.8$ & $466.3$ &  $1.9$ & $1564$ & $4.2$ &  $9.9$ & & Fig.~\ref{kmass}(b) \\
                    &            & $8$ & & $843.8$ & $291.1$ & $432.8$ & $567.5$ &    $-$ &    $-$ & $3.3$ &  $7.7$ & & Fig.~\ref{kmass}(c) \\
\hline
({\em S2}) Order-disorder & $m_{B2}/m_{B1}=$ & $2$ & & $381.6$ & $20.4$ & $38.7$ & $201.0$ & $-$ & $-$ & $5.7$ & $33.2$ & & Fig.~\ref{kodis}(a) \\
                    &                  & $4$ & & $381.6$ &  $9.9$ & $19.3$ & $195.8$ & $-$ & $-$ & $4.5$ & $14.3$ & & Fig.~\ref{kodis}(b) \\
                    &                  & $8$ & & $381.6$ &  $7.7$ & $15.1$ & $194.6$ & $-$ & $-$ & $4.0$ &  $8.2$ & & Fig.~\ref{kodis}(c) \\
\hline
({\em S3}) Weak interface & $\epsilon_{AB}=$ & $0.5$ & & $587.3$ & $587.3$ & $-$ & $-$ & $-$ & $-$ & $10.6$ & $-$ & & Fig.~\ref{klj}(a) \\
                     &                  & $0.1$ & & $587.3$ & $587.3$ & $-$ & $-$ & $-$ & $-$ & $10.6$ & $-$ & & Fig.~\ref{klj}(b) \\
\hline
\hline
\end{tabular}
\end{table*}

Although the above works have demonstrated very low values of $\kappa$ in superlattice systems, we note that these have not been systematically compared to the values assumed in the glasses with {\em exactly} the same chemical composition. Also, a general framework to rationalize in a coherent single picture all these observations is, to the best of our knowledge, still lacking. 

In this work, we address these two issues.  Building on the comparison of the superlattice with the corresponding amorphous structure, we clarify the mechanisms allowing for ultra-low thermal conductivity in the former. We have studied by computer simulation a numerical model that allows one to exactly compare ordered and disordered systems with identical chemical composition and access detailed information on the entire normal modes spectrum, providing, as a consequence, a complete understanding of the heat transfer process. As the lifetime of heat carriers is already minimum in glasses~\cite{Mizuno2_2013,Mizuno2_2014}, we demonstrate that the key to even lower thermal conductivities is to suppress  their propagation across the interfaces between the constituent layers. 

More in details, we have focused on three distinct design principles for superlattices, mimicking similar configurations actually employed in experiments. These are based on the face-centered-cubic (FCC) lattice structure, and are composed of: ({\em S1}) two intercalated crystalline layers formed by sphere particles with different masses; ({\em S2}) ordered crystalline layers intercalated to mass-disordered alloy layers; and ({\em S3}) identical crystalline layers with modified (weakened) interactions across the interfaces (see the Methods section and Table~\ref{case}). We show that a large mass difference between layers ({\em S1}) and weakened interactions between layers ({\em S3}) efficiently  obstruct the  propagation of phonons, resulting in a very large reduction of the superlattice thermal conductivity, even below the values pertaining to the glass phases with identical composition. Based on our results, we conclude with a discussion of the optimal strategy to follow towards {\em very} low thermal conductivity materials.

In Fig.~\ref{schematic} we show a schematic illustration of a superlattice composed by two intercalated layers, $A$ and $B$, both of thickness $w/2$. The competition between two length scales,  the repetition period of the superlattice, $w$, and the mean free path of the superlattice phonons, $\ell$, determines the coherent or incoherent character of phonon transport, as described in~\cite{Simkin_2000,Yang_2003,Garg_2013} and demonstrated by numerical simulations~\cite{Chen_2005,Kawamura_2007,Yang_2008} and recent experiments~\cite{Ravichandran_2014}. 

For $w>\ell$, the {\em incoherent} phonon transport is independent in the different layers, and phonons can be effectively treated as particles. In this case, the Boltzmann transport equation applies~\cite{Chen_1997,Chen_1998}, and the particle-like phonons are scattered within the layers (internal resistance) and at the interfaces (interfacial resistance)~\cite{Kim2_2000,Lampin_2012}. The thermal conductivity in the cross-plane direction can be written as
\begin{equation}
\kappa_\text{CP}=\frac{2}{\kappa_A^{-1}+\kappa_B^{-1}+4Rw^{-1}}=\kappa_{\text{CP}}^{\infty}\left(\frac{1}{1+ \ell_K w^{-1}} \right),
\label{equdiffusecp}
\end{equation}
where
\begin{equation}
\kappa_{\text{CP}}^{\infty}=\lim_{w\rightarrow\infty}\kappa_\text{CP}=\frac{2\kappa_A \kappa_B}{ \kappa_A + \kappa_B }. 
\end{equation}
Here, $\kappa_A$ and $\kappa_B$ are the thermal conductivities of materials $A$ and $B$, and $\ell_K=2R \kappa_{\text{CP}}^{\infty}$ is the Kapitza length~\cite{Nan_1998,BARRAT_2003}. $R$ is the interfacial resistance, which exists even at a perfect interface and depends on the nature of the contacting materials (e.g., crystal-crystal, crystal-glass)~\cite{Kim2_2000,Lampin_2012}. For $w<\ell_K$ ($w>\ell_K$), the interfacial resistance is relatively large (small) compared to the internal resistance. Both $\kappa_\text{CP}$ and $\kappa_\text{IP}$ (the in-plane thermal conductivity) increase with $w$, due to the decrease of the interfacial resistance density~\cite{Chen_1997,Chen_1998}. In the diffuse limit $w \rightarrow \infty$, where the interfacial resistance can be neglected, $\kappa_\text{CP}$ and $\kappa_\text{IP}$ have the upper bounds $\kappa_{\text{CP}}^{\infty}$ and $\kappa_{\text{IP}}^{\infty}=(\kappa_A + \kappa_B)/2$, respectively.

When $w<\ell$, phonon transport is {\em coherent}~\cite{Simkin_2000,Yang_2003,Garg_2013,Chen_2005,Kawamura_2007,Yang_2008,Ravichandran_2014}, and the wave nature of phonons cannot be neglected. In this regime, $\kappa_\text{CP}$ decreases with increasing $w$, in contrast with the incoherent case. The reduction of $\kappa_\text{CP}$ is explained with the emergence of a band gap at the Brillouin zone boundary, due to band-folding~\cite{Tamura_1988,Mizuno_1992}: increasing $w$ augments the frequency gap in the dispersion relation. This, in turn, decreases the average group velocity $v$ of phonons, finally reducing $\kappa_\text{CP}$. Mini-umklapp processes~\cite{Ren_1982}, occurring at the mini-Brillouin zone, also contribute to the reduction of $\kappa_\text{CP}$. At the crossover length $w \sim \ell$, between the incoherent and the coherent transport regimes, $\kappa_\text{CP}$ assumes a minimum value when plotted against $w$~\cite{Simkin_2000,Yang_2003,Garg_2013,Chen_2005,Kawamura_2007,Yang_2008,Ravichandran_2014}. We have encountered this situation in the case of superlattice {\em S1}, as we will see below.

Details of the structure of the interface between layers are also known to significantly affect phonon transport~\cite{Daly2_2002,Imamura_2003,Daly_2003,Landry_2009,Huberman_2013,Termentzidis_2009,Termentzidis_2011,Hsieh_2011,Shen_2011,Losego_2012,Wei_2013}. It has been reported that interfacial roughness~\cite{Daly2_2002,Imamura_2003,Daly_2003} or mixing~\cite{Landry_2009,Huberman_2013} reduce both $\kappa_\text{CP}$ and $\kappa_\text{IP}$, and can even suppress the coherent nature of phonons, with $\kappa_{\text{CP}(\text{IP})}$ increasing monotonously at any $w$. The interface topology is also an important factor to determine the phonon transport~\cite{Termentzidis_2009,Termentzidis_2011}. While we will not address precisely this situation in detail here, the superlattice {\em S2} of our study bears some similarities with it.

Finally, the stiffness of interfacial bondings, which can be controlled by applying pressure~\cite{Hsieh_2011,Shen_2011} or tuning chemical bonding~\cite{Losego_2012}, has significant effects on  heat transport features, which will be demonstrated by the study of the {\em S3} superlattice.
\begin{figure}[t]
\centering
\includegraphics[width=0.425\textwidth]{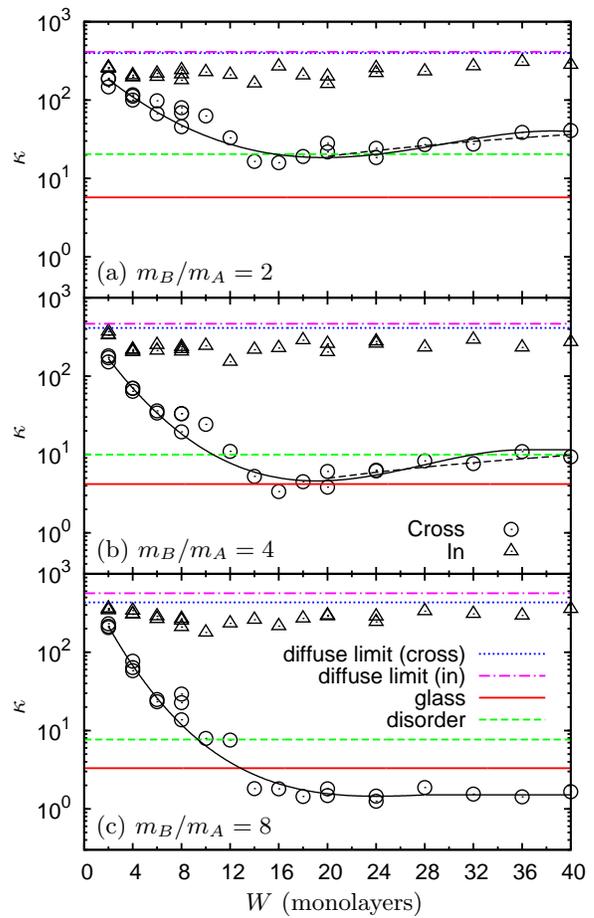}
\caption{
{\bf Thermal conductivity in superlattice {\em S1} composed of two intercalated crystalline layers with different masses.} The cross-plane, $\kappa_\text{CP}$, and in-plane, $\kappa_\text{IP}$, components of thermal conductivity are plotted as functions of the repetition period $w$. The ratio $m_B/m_A$ of the masses in layers $A$ and $B$ is $2$ in panel (a), $4$ in (b), and $8$ in (c). The values $\kappa_\text{CP}^\infty$ and $\kappa_\text{IP}^\infty$ of the diffuse limits ($w\rightarrow \infty$), as well as those in the glass and the disordered alloy with the same constituent species are indicated by the horizontal lines. In panels (a) and (b) we also show (dashed black lines), the prediction of Eq.~(\ref{equdiffusecp}) for $\kappa_\text{CP}$ in the incoherent regime, $w>20$, with the values of $R$ and $\ell_K$ included in Table~\ref{case}. The solid curve interpolating the $\kappa_\text{CP}$ data points in the entire $w$-range is a guide for eye. For some values of $w$, multiple data points are shown, calculated by using different system sizes in order to exclude the presence of finite system size issues (see the {\bf Methods} section for details on this point).
} 
\label{kmass}
\end{figure}
\begin{figure*}[t]
\centering
\includegraphics[width=0.935\textwidth]{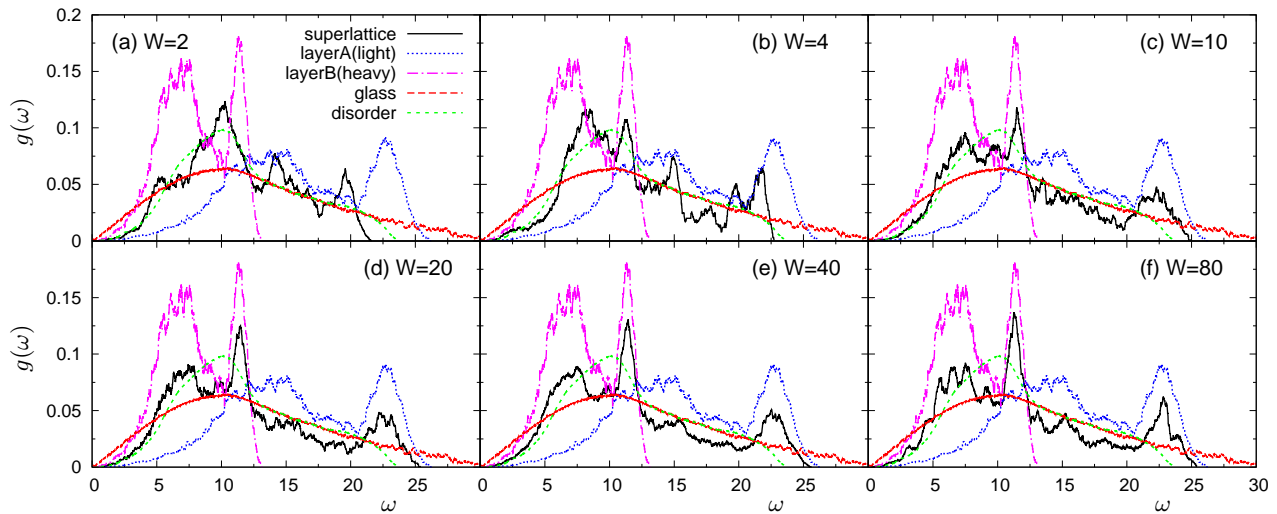}
\caption{
{\bf Vibrational density of states in superlattice {\em S1}.} Vibrational density of states data for a mass ratio $m_B/m_A=4$, with $m_A=0.4$ and $m_B=1.6$. In panels (a)-(f) we show the data corresponding to the repetitions period values $w=2, 4, 10, 20, 40, 80$. For  comparison, we also plot $g_{A(B)}(\omega)$ for the homogeneous bulk crystal composed by light (heavy) $m_{A(B)}$ masses only, together with the data for the glass and the disordered alloy formed by the same constituent species.
} 
\label{vmass}
\end{figure*}

\section{Results}
In Table~\ref{case}, we present the details of the three superlattice systems studied in this work, with values of the important quantities: $\kappa_A$ and $\kappa_B$ are the thermal conductivities of layers $A$ and $B$, respectively; $\kappa_{\text{CP}}^{\infty}$ and $\kappa_{\text{IP}}^{\infty}$ are the cross- and in-plane diffuse limits of $\kappa_{\text{CP}}$ and $\kappa_{\text{IP}}$; $R$ is the interfacial resistance, $\ell_K$ the Kapitza length; $\kappa_{\text{glass}}$ and $\kappa_{\text{disorder}}$ are the thermal conductivities of the glass and disordered alloy with exactly the same composition as  the indicated superlattice. Thermal conductivities have been estimated by molecular-dynamics (MD) simulation and the Green-Kubo formulation~\cite{McGaughey,Landry_2008}. The number density and the temperature are fixed at $\hat{\rho}=1.015$ (the corresponding crystal lattice constant is $a=1.58$) and $T=10^{-2}$, respectively. Vibrational states were also characterized by using a standard normal-modes analysis~\cite{Ashcroft}. Details about the systems and the methods used for the simulation production runs and analysis are given in the {\bf Methods} section.
\\

\noindent
{\bf S1. Superlattice composed of two intercalated crystalline layers with different masses.} In Fig.~\ref{kmass} we show the thermal conductivities, $\kappa_\text{CP}$ and $\kappa_\text{IP}$ (symbols), as functions of the replication period, $w$, for the layers mass ratios $m_B/m_A=2,\ 4$, and $8$. The values of the diffuse limits $\kappa_\text{CP}^\infty$ and $\kappa_\text{IP}^\infty$ as well as those of the glass and the disordered alloy constituted by the same species (see Table~\ref{case}) are also shown as lines. As expected, the relation $\sqrt{m_A} \kappa_A= \sqrt{m_B} \kappa_B$ holds for the pure materials. In the studied $w$-range, $w=2$ to $40$ (monolayers), the in-plane value $\kappa_\text{IP}$ shows a very weak dependence on $w$, as was observed for superlattices with perfect interfaces in Refs.~\cite{Kawamura_2007,Daly_2003}. The value of $\kappa_\text{IP}$ is close to, although lower than, $\kappa_\text{IP}^\infty$, indicating that slight in-plane phonon scattering at the interface is still active.
\begin{figure*}[t]
\centering
\includegraphics[width=0.935\textwidth]{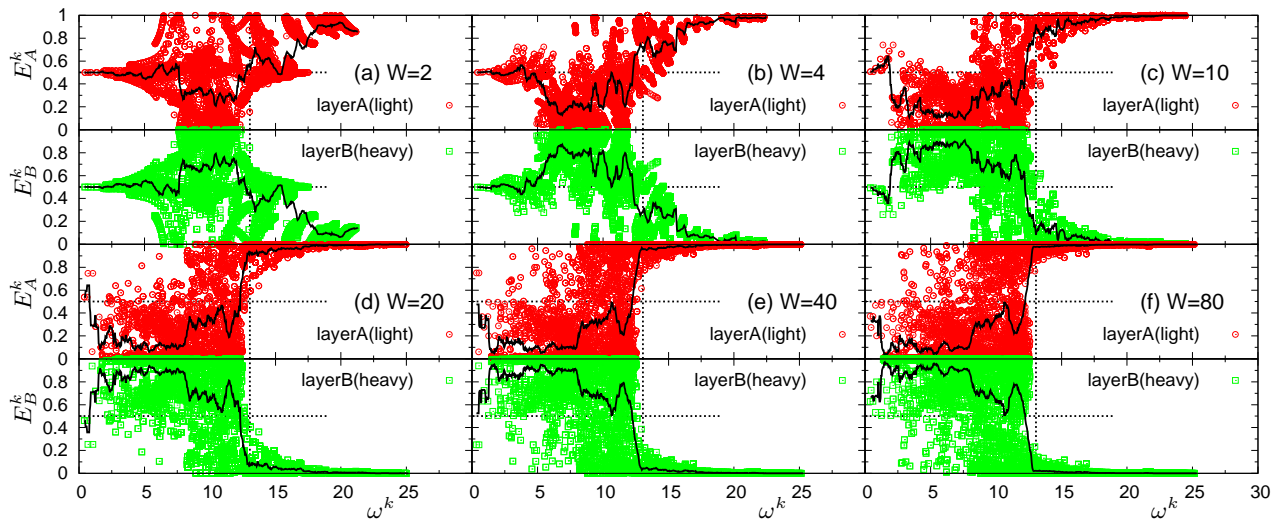}
\caption{
{\bf Vibrational amplitudes of normal modes in superlattice {\em S1}.} Vibrational amplitudes of eigenvectors, $E_A^k$ and $E_B^k$, in layers $A$ (light) and $B$ (heavy) for all normal modes $k$, plotted as functions of the corresponding eigenfrequency $\omega^k$. $E_A^k$ and $E_B^k$ are defined in Eq.~(\ref{amp}). The mass ratio of the two layers is $m_B/m_A=4$, and the repetitions period values are $w=2, 4, 10, 20, 40, 80$ in panels (a)-(f). The solid line represents the average values $\left< E_A^k \right>$ and $\left< E_B^k \right>$ calculated in bins of the form $\omega^k\pm\delta\omega^k/2$, with $\delta \omega^k=0.5$. The horizontal dotted lines represent the threshold value $E_A^k=E_B^k=0.5$, the vertical lines indicate $\omega = \omega^\text{max}_B\simeq 13$, corresponding to the high frequency edge of $g_B(\omega)$.
} \label{emass}
\end{figure*}

More interestingly, as $w$ increases, the cross-plane value $\kappa_\text{CP}$ decreases steeply, reaches a minimum value at $w^* \simeq 20$, and next increases mildly at larger $w$. This $w$-dependence is consistent with previous predictions~\cite{Simkin_2000,Yang_2003,Garg_2013,Chen_2005,Kawamura_2007,Yang_2008,Ravichandran_2014}, and corresponds to the crossover at $w^*$ from coherent to incoherent phonon transport. In the incoherent regime, $w > 20$, from Eq.~(\ref{equdiffusecp}) and the data of $\kappa_\text{CP}$ (dashed line in Fig.~\ref{kmass}) we can extract the values of the interfacial resistance, $R$, and the Kapitza length, $\ell_K$, which are presented in Table~\ref{case}. Note that for $m_B/m_A=8$ (Fig.~\ref{kmass}(c)), we do not observe a clear thermal conductivity minimum. More precisely, even at the largest value $w=40$, $\kappa_\text{CP}$ is still orders of magnitude lower than $\kappa_\text{CP}^\infty$, indicating that the interfacial resistance $R$ results in a strong reduction of $\kappa_\text{CP}$ in this range of $w$. Equivalently, the Kapitza length $ \ell_K$ is significantly larger than the maximum period $w=40$. The data shown in Fig.~\ref{kmass} demonstrate that $\kappa_\text{CP}$ can be indeed lowered below the disordered alloy limit for $m_B/m_A=2$, and even below the glass limit for higher mass heterogeneities, $m_B/m_A=4$ and $8$. These results are consistent with the experimental work of Ref.~\cite{Costescu_2004}, and demonstrate that the interface formed between dissimilar materials effectively reduces $\kappa_\text{CP}$. It is also worth noting that the thermal conductivity tensor is very strongly anisotropic in this case, with $\kappa_\text{CP} \ll \kappa_\text{IP}$.

The vibrational modes of the structure, i.e., the superlattice phonons, are key to understand the above behaviour of thermal conductivity. In Fig.~\ref{vmass} we show the vibrational density of states (vDOS), $g(\omega)$, for $m_B/m_A=4$ and $w=2$ to $80$. $g_A(\omega)$ and $g_B(\omega)$ of the bulk crystals of type $A$ and $B$. The vDOS of the glass and of the disordered alloy   are also shown for comparison. Note that $g_A(\sqrt{m_A}\omega)/\sqrt{m_A} = g_B(\sqrt{m_B}\omega)/\sqrt{m_B}$. At small $w=2$, $g(\omega)$ of the superlattice roughly follows that of the disordered alloy, implying that the vibrational states in the two layers are strongly mixed. In this situation, phonons are able to propagate in both the cross- and in-plane directions. On the other hand, as $w$ increases, $g(\omega)$ generates features increasingly similar to those identifying $g_A(\omega)$ and $g_B(\omega)$, separately. In particular, in the low-$\omega$ region  $g(\omega)$ follows $g_B(\omega)$ (the heavy crystal $B$), whereas $g_A(\omega)$ (the light crystal $A$) controls $g(\omega)$ in the high-$\omega$ region. This result indicates that different parts of the vibrational spectrum are active in the two layers, with high(low)-$\omega$ modes preferentially excited in the light (heavy) layer $A$ ($B$). In this situation, phonon propagation is largely obstructed in the cross-plane direction, leading to the observed large reduction of $\kappa_\text{CP}$. We remark that phonons propagate in the in-plane direction with few constraints, as shown by the large value of $\kappa_\text{IP}$ close to $\kappa_\text{IP}^\infty$. This implies that phonons, whose propagations are blocked in the cross-plane direction, are actually specularly reflected at the interface and confined in the in-plane direction.

The separation of the vibrational states found in the $g(\omega)$ becomes more clear when considering the vibrational amplitudes associated with the eigenstates $k$. In Fig.~\ref{emass} we show the vibrational amplitudes, $E_A^k$ and $E_B^k$ (Eq.~(\ref{amp})), in the two layers $A$ and $B$ for each mode $k$, together with the binned average values (solid lines). Based on the relations $E^A_k+E^B_k = 1$ and $0 \le E_A^k, E_B^k \le 1$, we can define a relative degree of excitation of particles in the two layers, by the threshold value $0.5$: large excitations correspond to $E_{A,B}^k \ge 0.5$, small excitations to $E_{A,B}^k < 0.5$. If $E_A^k = E_B^k= 0.5$, particle vibrations in both layers are of the same degree and correlated. 

At small $w=2, 4$ we find, particularly in the low-$\omega$ region, a large fraction of vibrational states with $E_A^k \simeq E_B^k \simeq 0.5$. As $w$ increases, in the high frequency region $\omega > \omega^\text{max}_B$, where $\omega^\text{max}_B \simeq 13$ is the high-frequency boundary in $g_B(\omega)$, only particles in the light layer $A$ vibrate ($E_A^k \simeq 1$), whereas those in the heavy layer $B$ are almost immobile, as indicated by $E_B^k \simeq 0$. In this $\omega$-region, phonon propagation in the cross-plane direction is therefore almost completely suppressed. On the other hand, for $\omega < \omega^\text{max}_B$, particles pertaining to the heavy layer $B$ show large vibrational amplitudes ($E_B^k > 0.5$), while vibrations in layer $A$ tend to be small ($E_A^k < 0.5$). More in details, for $w \ge 20$, we see that the  averaged amplitudes are much larger in the $B$ layer ($\left< E_B^k \right> > 0.8$) than in the $A$ layer ($\left< E_A^k \right> < 0.2$) in the $2 < \omega < 7.5$ range. Contrary to the case of $\omega > \omega^\text{max}_B$, however, a significant number of modes are excited in both layers $A$ and $B$, even with $E_A^k\simeq E_B^k \simeq 0.5$. We therefore conclude that, for $\omega < \omega^\text{max}_B$, some phonons still propagate in the cross-plane direction, contributing to $\kappa_\text{CP}$.
\begin{figure}[t]
\centering
\includegraphics[width=0.425\textwidth]{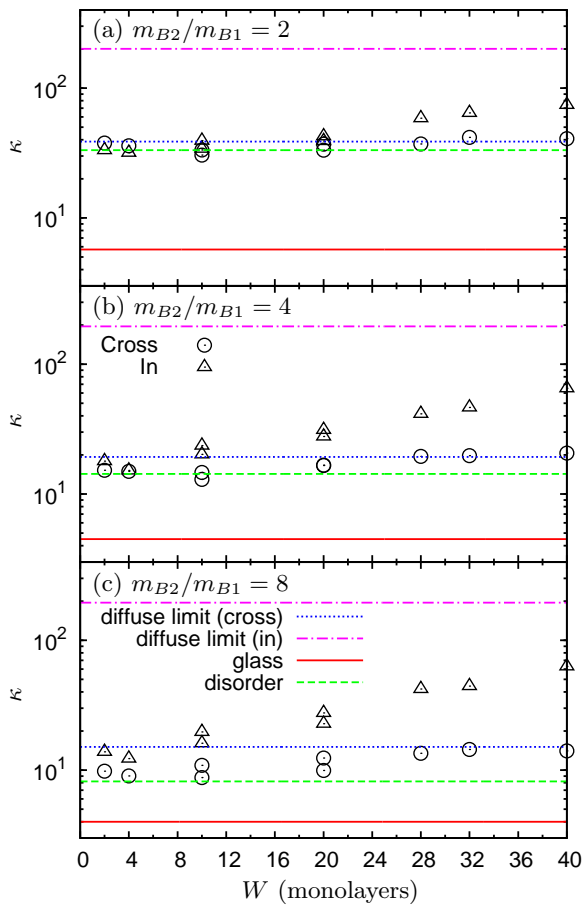}
\caption{
{\bf Thermal conductivity in superlattice {\em S2} composed of ordered crystalline layers intercalated with mass disordered  layers.} The two components $\kappa_\text{CP}$ and $\kappa_\text{IP}$ are plotted as functions of $w$. The mass ratio of the disordered alloy layer $m_{B2}/m_{B1}$ is (a) $2$, (b) $4$, and (c) $8$. The values $\kappa_\text{CP}^\infty$ and $\kappa_\text{IP}^\infty$ of the diffuse limits, together with those in the glass and the disordered alloy are indicated by lines.
} 
\label{kodis}
\end{figure}
\begin{figure}[b]
\centering
\includegraphics[width=0.425\textwidth]{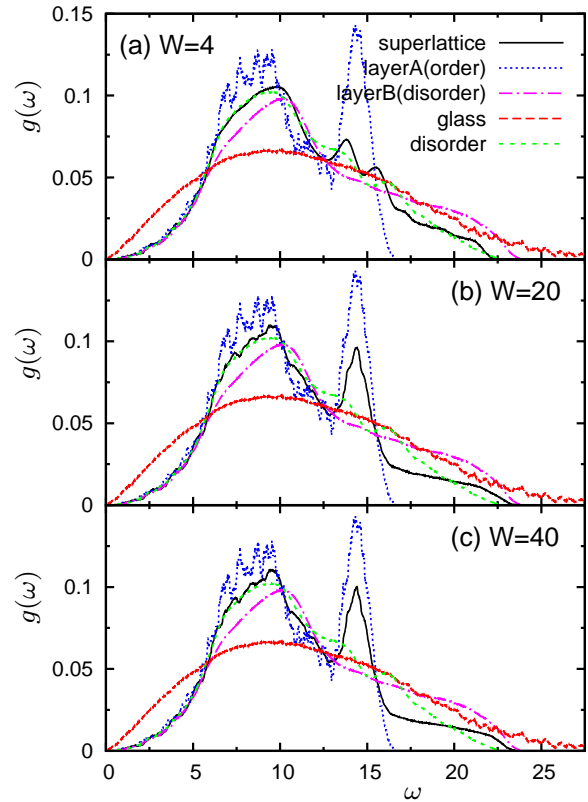}
\caption{
{\bf Vibrational density of states in superlattice {\em S2}.} We show our results for the mass ratio of the disordered alloy layer $m_{B2}/m_{B1}=4$, with $m_{B1}=0.4$ and $m_{B2}=1.6$. The period $w$ is $4$, $20$, and $40$ for (a), (b), and (c), respectively. For comparison, we plot $g_A(\omega)$ of the bulk crystal formed by particles of mass $m_A=1$ (layer $A$), $g_B(\omega)$ of the disordered alloy with masses $m_{B1}=0.4$ and $m_{B2}=1.6$ (layer $B$), and the vDOS of the glass and the disordered alloy formed by the same constituent species.
} 
\label{vodis}
\end{figure}

We note that our observation of the vibrational separation in both the vDOS and vibrational amplitudes is consistent with results reported previously~\cite{Yang_2008,Huberman_2013,Yang_2012}. Indeed, the simulation work of Ref.~\cite{Yang_2008} reported a separation in the vDOS of the Si isotopic-superlattice (${}^{28}\text{Si}$-${}^{42}\text{Si}$ superlattice). A recent simulation work~\cite{Huberman_2013} focused on partial inverse participation ratios in a superlattice similar to the one considered here, reporting vibrational modes separation between layers. Ref.~\cite{Yang_2012} attributed the reduction of thermal conductivity to a mechanism described as phonon localization, which we consider to be essentially the same phenomenon as the vibrational separation described here.

We believe that this concept of vibrational separation is a simple and accurate framework to rationalize the behaviour of thermal conductivity in superlattices. In particular, it provides a complete characterization of the minimum in the $w$-dependence of $\kappa_\text{CP}$. Indeed, in the range $w=2$ to $20$ identifying the coherent regime, the vibrational separation hinders the coherent phonon propagation in the cross-plane direction, leading to the large reduction of $\kappa_\text{CP}$. In contrast, in-plane phonon propagation is very mildly affected by the vibrational separation and, therefore, $\kappa_\text{IP}$ keeps high values. Also, by considering $\left< E_A^k \right>$ and $\left< E_B^k \right>$ (solid lines), we conclude that the separation saturates to its maximum level at $w \simeq 20$. Upon further increase $w>20$, although averaged values show no significant changes, we recognize an increasing fraction of modes with $E_A^k>0.5$ and $E_B^k<0.5$ for $\omega < \omega^\text{max}_B$ (panels (e) $w=40$ and (f) $w=80$ in Fig.~\ref{emass}). This observation indicates that the separation tendency for modes with $E_A^k<0.5$ and $E_B^k>0.5$ becomes weaker, i.e., the correlation of vibrational features in the two layers decreases, which corresponds exactly to the incoherent transport picture, and leads to the increase of $\kappa_\text{CP}$. Although transport becomes completely incoherent only for values of $w$ of the order of the Kapitza length (note that $\ell_z \simeq 1600$ for $m_B/m_A=4$), this feature appears as soon as the vibrational separation is saturated, at the crossover point $w^* \simeq 20$. Thus, the saturation point of the vibrational separation identifies the minimum value of $\kappa_\text{CP}$, which can be indeed below the glass limit.
\\

\noindent
{\bf S2. Superlattice composed of intercalated ordered crystalline layers and  mass disordered layers.} This system consists of three components, with masses $m_A=1$ in the crystalline layer $A$, and $m_{B1}$ and $m_{B2}$ in the disordered alloy layer $B$. In Fig.~\ref{kodis}, we plot $\kappa_\text{CP}$ and $\kappa_\text{IP}$ for the mass ratios of the layer $B$, $m_{B2}/m_{B1}=2$, $4$, and $8$. At small $w \le 4$, the values of both $\kappa_\text{CP}$ and $\kappa_\text{IP}$ are very close to those of the disordered bulk alloy formed by the same particles. As $w$ increases, $\kappa_\text{IP}$ increases gradually toward $\kappa_\text{IP}^\infty$. This increase is controlled by the development of in-plane phonon propagation in the ordered crystalline layer $A$. Indeed, the $g(\omega)$ of the superlattice, shown in Fig.~\ref{vodis}, roughly follows that of the disordered bulk alloy at small $w=4$, whereas at large $w=20, 40$ it is dominated by $g_A(\omega)$. In particular, the longitudinal peak around $\omega \simeq 14.5$ becomes clear, corresponding to that of the crystalline layer $A$.
\begin{figure}[t]
\centering
\includegraphics[width=0.425\textwidth]{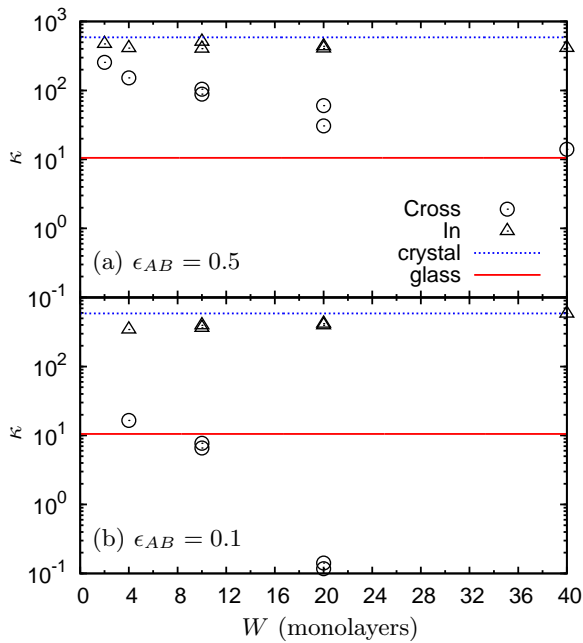}
\caption{
{\bf Thermal conductivity in superlattice {\em S3} composed of identical crystalline layers with weakened interface.} Thermal conductivities $\kappa_\text{CP}$ and $\kappa_\text{IP}$ are plotted as functions of $w$. The interface interaction $\epsilon_{AB}$ is $0.5$ in (a) and $0.1$ in (b). We also show, by the horizontal lines, the thermal conductivities of the corresponding one-component homogeneous bulk crystal and glass with unmodified interactions.
} 
\label{klj}
\end{figure}

The cross-plane value $\kappa_\text{CP}$ also increases with $w$, but reaches the limit value $\kappa_\text{CP}^\infty$ already at $w\sim 20$. Since $\kappa_B$ of the disordered alloy layer $B$ is low (see Table~\ref{case}), $\kappa_\text{CP}^\infty$ remains low, typically less than twice the disordered alloy value. As a result, the variation of $\kappa_\text{CP}$ with $w$ is small. This result indicates that scattering in the disordered alloy layer $B$ dominates the thermal conduction in the cross-plane direction. Both experimental work~\cite{Li_2003} on Si(crystal)-SiGe(disordered alloy) nanowires and numerical simulations~\cite{Landry_2009} have reported similar observations. We also note that the coherent nature of the superlattice phonons in the cross-plane direction, which we observed in the {\em S1} system, breaks down in {\em S2}. This is essentially equivalent to previous findings that disorder in interfacial roughness~\cite{Daly2_2002,Imamura_2003,Daly_2003}, or interfacial species mixing~\cite{Landry_2009,Huberman_2013} destroy the coherent features of vibrational excitations present in the investigated superlattices. In addition, the thermal conductivity tensor becomes increasingly anisotropic at larger $w$ due to the increase of $\kappa_\text{IP}$, showing a behaviour different than that observed in {\em S1}. As a consequence of these features, in superlattices of type {\em S2} the variability of the cross-plane heat transport is strongly bounded, and the minimum limit of $\kappa_\text{CP}$ just corresponds to the disordered alloy limit, i.e., $\kappa_\text{CP}$ cannot be reduced below the glass limit.
\\
\begin{figure}[b]
\centering
\includegraphics[width=0.425\textwidth]{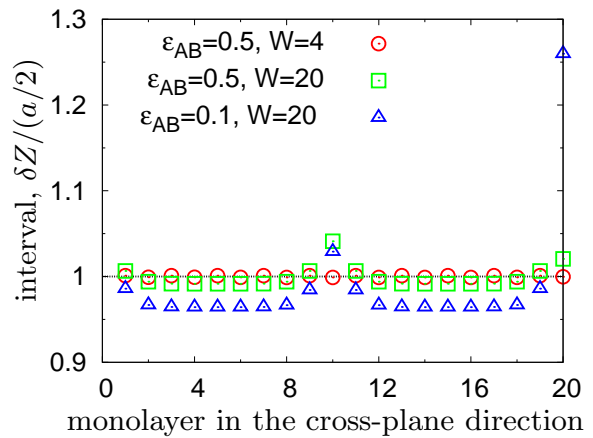}
\caption{
{\bf Distance between adjacent monolayers in superlattice {\em S3}.} The average cross-plane distance $\delta z$ between adjacent crystalline planes plotted for each monolayer, identified by the corresponding order index. We present the value of $\delta z$ normalized to $a/2$, the horizontal line $\delta z/(a/2) = 1$ therefore indicates the value in the perfect crystalline lattice. The displacements observed in the cases $w=20$ are discussed in the main text.
} 
\label{dlj}
\end{figure}
\begin{figure*}[t]
\centering
\includegraphics[width=0.935\textwidth]{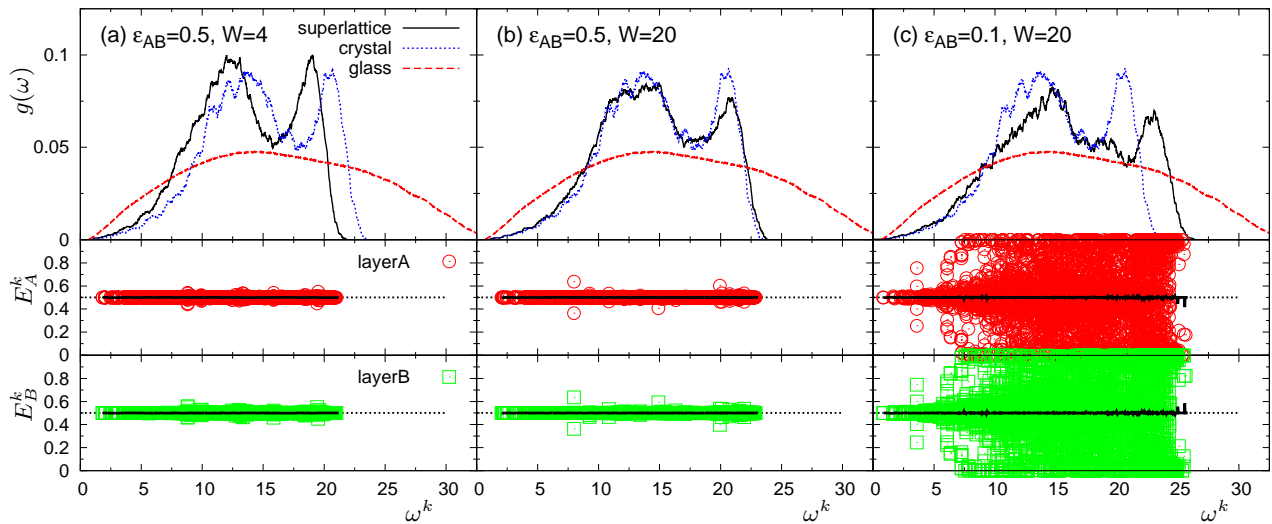}
\caption{
{\bf Vibrational density of states and vibrational amplitudes in superlattice {\em S3}.} We report data corresponding to the indicated values of the the interfacial interaction energy $\epsilon_{AB}$ and the repetition period $w$: (a) $\epsilon_{AB}=0.5$, $w=4$, (b) $\epsilon_{AB}=0.5$, $w=20$, and (c) $\epsilon_{AB}=0.1$, $w=20$. In panels at the top, we show the vDOS $g(\omega)$ for the superlattices of type {\em S3}, together those of the corresponding one-component homogeneous crystal and glass with unmodified interactions. In panels at the bottom we show the vibrational amplitudes, $E_A^k$ and $E_B^k$, in layers $A$ and $B$, plotted as functions of the eigenfrequency $\omega^k$. The solid line represents the average values $\left< E_A^k \right>$ and $\left< E_B^k \right>$, calculated in bins of the form $\omega^k\pm\delta\omega^k/2$, with $\delta \omega^k=0.5$. The horizontal dotted lines indicate $E_A^k=E_B^k=0.5$.
} 
\label{velj}
\end{figure*}

\noindent
{\bf S3. Superlattice composed by identical crystalline layers separated by weakly interacting interfaces.} In Fig.~\ref{klj} we show the $w$-dependences of $\kappa_\text{CP}$ and $\kappa_\text{IP}$ for the case where the energy scale associated to particles interactions across the interfaces ($\epsilon_{AB}$) are lowered compared to those intra-layers, with $\epsilon_{AB}=0.5$ and $0.1$ in the two panels. In the figure, we also plot as lines the data for the corresponding one-component crystal and glass with unmodified interactions. The in-plane value $\kappa_\text{IP}$ is almost independent of $w$, and is very close to the value pertaining to the crystal. In contrast, $\kappa_\text{CP}$ decreases monotonically by increasing $w$, and especially in the weaker case $\epsilon_{AB}=0.1$, the observed reduction of $\kappa_\text{CP}$ is dramatic. At $w=10$, $\kappa_\text{CP}$ equals the value obtained for the glassy sample, and it is  almost two orders of magnitude lower than this value at $w=20$. This extremely low $\kappa_\text{CP}$ is consistent with previous experimental work~\cite{Chiritescu_2007}.

Some insight about the origin of this observation comes from the data shown in Fig.~\ref{dlj}, where we display the average cross-plane distance $\delta z$ between adjacent crystalline planes (monolayers), normalized to the value in the perfect lattice, $a/2$. For $\epsilon_{AB}=0.5$ and $w=4$, the system keeps the perfect lattice structure, with $\delta z \equiv a/2$ for all monolayers. In contrast, as $\epsilon_{AB}$ decreases and for a large value $w=20$, $\delta z$ becomes substantially larger than $a/2$ at the interfaces, which therefore assumes a local density lower than the average. At the same time, slightly reduced $\delta z$ are also observed for the other intra-monolayers, leading to an increase of the local density compared to the average. This heterogeneity hinders energy propagation across the interface and, as a result, phonons are specularly reflected and confined in the in-plane direction. We remark that in the cases with $w=20$, the values of $\delta z$ at the interfaces located at $w/2=10$ and $w=20$ are different, with a large discrepancy for $\epsilon_{AB}=0.1$. We rationalize this behaviour by observing that, during the preparation stage of the sample, the applied selective weakening of the interactions destabilizes the global equilibrium of the superlattice, with a concentration of mechanical stress close to the interfaces. Lattice planes far from the boundaries easily recover mechanical equilibrium by coherently reducing their mutual distance. In contrast, particles in monolayers adjacent to the interfaces move both out-of-plane and in-plane, to optimize the local effective spring constants. The optimal solution found depends in general on the details of the local environment, explaining the observed discrepancy in $\delta z$ at different interfaces.   

The behaviour of $\kappa_\text{CP}$ can be further elucidated by inspection of the main features of the vibrational spectrum. In Fig.~\ref{velj} we plot the $g(\omega)$ of superlattice {\em S3}, together with the vibrational amplitudes $E_A^k$ and $E_B^k$. The $g(\omega)$ shows transverse and longitudinal phonon branches for all cases, similar to the homogeneous bulk crystal. As $w$ increases $g(\omega)$ deforms, following the appearance of an increasing fraction of modes at increasing higher frequencies. This behaviour is certainly correlated to the observation made above (see Fig.~\ref{dlj}) for $w=20$, that the distance between monolayers far from the interfaces becomes smaller than $a/2$. The consequent larger mass density makes higher the frequency of phonon modes of given wavelength, leading to the shift of $g(\omega)$ towards  higher frequencies. This global shift has as a consequence a mild increase of $\kappa_\text{IP}$ with $w$, as it is clear from Fig.~\ref{klj}(b) ($\epsilon_{AB} = 0.1$ case). Note that for $\epsilon_{AB}=0.5$ and $w=4$ (Fig.~\ref{velj}(a)), $g(\omega)$ shows an excess of lower-$\omega$ modes compared to those present in the one-component crystal, simply due to the weakened interactions at the interfaces.

We now focus on the vibrational amplitudes, $E_A^k$ and $E_B^k$ (Fig.~\ref{velj}, bottom panels). In the cases with $\epsilon_{AB}=0.5$ and $w=4$ and $20$, the particles in the two layers $A$ and $B$ show completely equivalent and correlated vibrations for the vast majority of the modes, as indicated by $E_A^k \equiv E_B^k \equiv 0.5$. This result implies that phonons indeed propagate across the weakened interfaces in the cross-plane direction, but they are also partially reflected at the interface, causing the observed reduction of $\kappa_{\text{CP}}$. The situation changes drastically in the case $\epsilon_{AB}=0.1$ and $w=20$, where the ultra-low value of $\kappa_\text{CP}$ can be reached. Except for the low-$\omega$ modes, $E^k_A$ and $E^k_B$ are symmetrically randomly distributed around the average values $\left< E_{A(B)}^k \right> \equiv 0.5$, indicating that particles in layers $A$ and $B$ vibrate independently, in an uncorrelated manner. As a consequence, a very large fraction of vibrational modes do not cross at all the interfaces, but rather undergo a perfect specular reflection. In this situation, heat is not transferred between two adjacent layers $A$ and $B$, leading to extremely low value of $\kappa_\text{CP}$, while keeping a high $\kappa_\text{IP}$. We conclude by noticing that although specular reflection was also observed in the system {\em S1}, the physical mechanism behind this phenomenon is different in the two cases: vibrational separation causes reflection in the former, whereas weakened interactions across the interface, with the resulting augmented spacing between the layers, completely block cross-plane phonon propagation in the latter.

\section{Discussion}
We have provided numerically, for the first time to our knowledge, a clear demonstration of very low thermal conductivities in superlattices, below the glassy limit of the corresponding amorphous structures. Blocking phonon propagation in ordered structures via interfaces design is the key principle. We have identified two possible strategies to achieve this goal: imposing a large mass heterogeneity in the intercalated layers (as in system {\em S1}) or degrading inter-layers interactions compared to those intra-layers (as in {\em S3}). We have found that in both cases phonons are specularly reflected at the interface and confined in the in-plane direction. This reduces the cross-plane thermal conductivity $\kappa_\text{CP}$ below the corresponding glass limit, while keeping the in-plane contribution $\kappa_\text{IP}$ close to the pure crystalline value. 

More specifically, in the case of mass mismatch ({\em S1}), propagation of phonons with high frequencies ($\omega > \omega_B^\text{max}$) is almost completely suppressed, whereas a fraction of low-frequency phonons ($\omega < \omega_B^\text{max}$) are still able to propagate across the interfaces, contributing to $\kappa_\text{CP}$ (Fig.~\ref{emass}(d)). Also, the minimum in thermal conductivity as a function of the repetition period $w$ (Fig.~\ref{kmass}) corresponds to a maximum in the vibrational separation between the layers of type $A$ and $B$. These therefore act as true filters in complementary regions of the vibrational spectrum, suppressing significantly phonons transport in the direction of the replication pattern. On the other hand, attenuated interactions across the interfaces ({\em S3}) are able to block phonons at almost all frequencies (see Fig.~\ref{velj}(c)), which results into extremely low values of $\kappa_\text{CP}$, even orders of magnitude lower than the corresponding glass limit (Fig.~\ref{klj}(b)). In this sense, directly modifying the interfaces seems to be the most effective strategy to obtain very low heat transfer. Note that this is a practically feasible route, since attenuated interfaces can be designed by exploiting materials with weak van der Waals forces among adjacent crystalline planes, as demonstrated in the case of $\text{WSe}_2$ sheets in Ref.~\cite{Chiritescu_2007}. Interfaces stiffness modification by controlling pressure~\cite{Hsieh_2011,Shen_2011} or chemical bonding~\cite{Losego_2012} are additional possible routes to directly tune the strength of interfaces.

Our data also suggest that intercalating disordered alloy layers in ordered crystalline layers ({\em S2}) is not effective in lowering $\kappa_\text{CP}$. Indeed, we have demonstrated that in this case disorder is not sufficient to block the propagation of vibrational excitations, even though it makes phonons lifetimes short. The intercalated disordered alloy layer dominates phonon transport in the entire superlattice, notwithstanding the presence of the crystalline layers. As a result, thermal conductivity is very similar to the one of the disordered alloy and is only marginally modified by modulation of the period $w$ (see Fig.~\ref{kodis}). Also, as suggested in previous works, disorder in the interfacial roughness~\cite{Daly2_2002,Imamura_2003,Daly_2003} or interfacial mixing~\cite{Landry_2009,Huberman_2013} seems to already dominate over phonon transport, and destroy the coherent nature of phonons.

In addition, as we understand from our analysis of vibrational amplitudes (Figs.~\ref{emass} and~\ref{velj}), it is much more problematic to block low-$\omega$ (long wavelength, $\lambda$) phonons propagation, than those with high-$\omega$ (short $\lambda$). This situation is similar to what has been observed in bulk glasses, where the long-$\lambda$ acoustic waves are not scattered by the disorder and can propagate over long distances by carrying heat energy~\cite{monaco2_2009,Mizuno_2014}. Therefore, blocking or efficiently scattering the long-$\lambda$ phonons is also a key factor to achieve very low thermal conductivities, as was pointed out in Ref.~\cite{Luckyanova_2012}. A possibility to realize this task is embedding in the targeted material objects featuring larger typical sizes, including nano-particles~\cite{Woochul_2006,Zhang_2014} or nano(quantum)-dots~\cite{Pernot_2010,Nika_2011}. Based on this strategy, very low thermal conductivity was achieved experimentally in a Si-Ge quantum-dot superlattice~\cite{Pernot_2010}, even below the amorphous Si value. The additional possibility of introducing large size defects by the porous structuring of materials has also been explored in a recent numerical work~\cite{Yang_2014}. Here, values of thermal conductivity $10^4$ times smaller than that of bulk Si were reached in Si phononic crystals with spherical pores.

In conclusion, we note that the three superlattice structures studied in the present work show totally different $w$-dependences of cross and in-plane thermal conductivities. Our results therefore not only contribute to a deeper comprehension of the physical mechanisms behind very-low thermal conductivity, they also provide insight for developing new design concepts for materials with controlled heat conduction behaviour.

\section{Methods}
{\bf System description.} In this Section we provide details on the numerical models we have used for the superlattices. The corresponding amorphous structures (glasses) and disordered alloys with exactly the same composition were also prepared, for the sake of comparison with superlattice phases. We have considered in all cases a 3-dimensional cubic box, of volume $V=L^3$ ($L$ being the linear box size), with periodic boundary conditions in all directions. In the superlattice and disordered alloy cases, particles were distributed on the FCC lattice sites. In the glass phases, they were frozen in topologically random positions following a rapid quench from the normal liquid phase below the glass transition temperature $T_g$, avoiding crystallization (see, for instance, Ref.~\cite{monaco2_2009} for details on the preparation of glasses). Particles, $i$ and $j$, interact via soft-sphere (SS) or Lennard-Jones (LJ) potentials:
\begin{equation}
\begin{aligned}
v_{SS}^{ij}(r)    &=  \epsilon^{ij} \left( \frac{\sigma^{ij}}{r} \right)^{12}, \\
v_{LJ}^{ij}(r) &= 4\epsilon^{ij} \left[ \left( \frac{\sigma^{ij}}{r} \right)^{12}- \left( \frac{\sigma^{ij}}{r} \right)^{6} \right],
\end{aligned}
\end{equation}
where $r$ is the distance between those two particles, and $\sigma^{ij}$ and $\epsilon^{ij}$ are the interparticle diameter and interaction energy scale, respectively. The potential is cut-off and shifted at $r_c=2.5 \sigma^{ij}$. Particle $i$ has mass $m^i$, and we have used $\sigma$, $\epsilon/k_B$ ($k_B$ is the Boltzmann constant), and $m$ as units of length, temperature, and mass. As a reference, for Argon $\sigma=3.4 \AA$, $\epsilon/k_B=120$~K, and $m=39.96$~a.m.u. We considered the number density $\hat{\rho}=N/V=1.015$, corresponding to a lattice constant $a=(4/\hat{\rho})^{1/3}=1.58$.

We prepared three superlattices, composed of intercalated FCC lattice layers, $A$ and $B$, both of thickness $w/2$, as schematically illustrated in Fig.~\ref{schematic}. The first superlattice ({\em S1}) consists of two crystalline layers formed by sphere particles with different masses, $m_A$ and $m_B$. We have considered mass ratios $m_B/m_A>1$, while keeping a constant average mass $(m_A+m_B)/2 = 1$. As an example, the case $m_B/m_A=4$ corresponds to $m_A=0.4$ and $m_B=1.6$. We have dubbed $A$ and $B$ as the light and heavy layers, respectively. Note that a mass ratio of $m_B/m_A=2.5$ corresponds to the case of the realistic Si-Ge superlattice. Except for the above mass difference in the different layers, all particles are characterized by the same properties. In particular, they interact via the SS potential $v_{SS}^{ij}(r)$, with $\sigma^{ij}=\epsilon^{ij}=1$.

The second superlattice ({\em S2}) is composed of an ordered crystalline layer $A$ intercalated to a disordered alloy layer $B$. $m_A=1$ in $A$, whereas in $B$ half of the particles have mass $m_{B1}$, $m_{B2}$ the others, and are randomly distributed on the lattice sites. Again, $m_{B1}$ and $m_{B2}$ are determined by the mass ratio $m_{B2}/m_{B1}>1$, keeping a constant average value $(m_{B1}+m_{B2})/2 = 1$. All particles in both layers interact via the SS potential $v_{SS}^{ij}(r)$ with $\sigma^{ij}=\epsilon^{ij}=1$.

The third superlattice ({\em S3}) is composed of identical crystalline layers $A$ and $B$, but the interactions among particles in different layers (i.e., across the interfaces) are modified (weakened) compared to those intra-layers. All particles have mass $m_A=m_B=1$, and interact via the LJ potential $v_{LJ}^{ij}(r)$, with  $\sigma^{ij}=\epsilon^{ij}=1$. The energy scale of interactions between particles pertaining to different layers are, however, reduced to $\epsilon^{ij}=\epsilon_{AB} <1$. 
\\

\noindent
{\bf MD simulation and the Green-Kubo method for the calculation of thermal conductivity.} In the present study, all simulations have been realized by using the large-scale, massively parallel molecular dynamics simulation tool LAMMPS~\cite{Plimpton_1995,lammps}. The systems were first equilibrated at relatively low temperature $T=10^{-2}$ by MD simulation in the $NVT$-ensemble. This choice was dictated by the need to reduce anharmonic effects, in order to primarily focus on the contribution of the structural features of the superlattices on thermal conductivity. We must note that our approach is classical, and does not take into account the quantum mechanisms active in the low-$T$ regime~\cite{kettel}. These effects have important implications, increasing the contribution to the thermal conductivity coming from low-$\omega$ vibrational excitations. At present, however, it is not obvious and still under debate how to effectively include quantum effects into a classical system~\cite{Turney_2009,Bedoya_2014}, and we have therefore chosen to stay within a fully classical approach. 

Following the equilibration stage, we performed the production runs in the $NVE$-ensemble. The Green-Kubo formulation~\cite{McGaughey,Landry_2008} was next applied to calculate the thermal conductivities, in the cross-plane and in-plane directions, respectively:
\begin{equation}
\begin{aligned}
\kappa_\text{CP} &= \frac{1}{VT^2} \int_{0}^\infty \left< {J}_z(t){J}_z(0) \right> dt,\\
\kappa_\text{IP} &= \frac{1}{2VT^2}\int_{0}^\infty \left< {J}_x(t){J}_x(0) + {J}_y(t){J}_y(0) \right> dt.
\end{aligned}
\end{equation}
Here, ${J}_{x,y}$, and ${J}_z$ are the heat currents in the in-plane ($x,y$) and cross-plane ($z$) directions, and $\left< \right>$ denotes the ensemble average. In the bulk glasses and disordered alloys, $\kappa_\text{CP}\simeq\kappa_\text{IP}$, i.e., heat conduction is isotropic, whereas in the superlattices, they are expected to assume different values~\cite{Yang_2002,Mavrokefalos_2007}. Landry \textit{et al.}~\cite{Landry_2008} have carefully confirmed the validity of the Green-Kubo method for the calculation of superlattices thermal conductivity, by comparison with the direct method based on non-equilibrium simulation. Also, in the Green-Kubo calculations, one must be attentive to finite system size effects~\cite{McGaughey,Landry_2008}. Indeed, long-wavelength phonons with $\lambda > L$ are excluded from the simulation box due to the finite value $L$ of the box size, which imposes important size effects on the numerical determination of $\kappa$. The box size therefore needs to be large enough to include a vibrational spectrum sufficient to establish an accurate description of anharmonic coupling (scattering) processes~\cite{McGaughey}. We note that the considered  $T=10^{-2}$ is low enough to substantially reduce anharmonic effects, but anharmonic couplings are still present. 

We can take care of finite size effects by increasing $L$ to values where $\kappa_\text{CP}$ and $\kappa_\text{IP}$ become $L$-independent. For the glass and disordered alloy thermal conductivities, we have confirmed that a system size $L=10a$ ($N=4,000$) is sufficiently large to obtain correct values of $\kappa_\text{CP} \simeq \kappa_\text{IP}$, without any size effect~\cite{Mizuno2_2013,Mizuno2_2014}. In the superlattice cases, the appropriate $L$ depends on the considered structure and the periodic repetition length $w$~\cite{Landry_2008}. More in details, we paid particular attention to the number $P$ of repetitions, defined from $L=Pw$, necessary to produce sufficient anharmonic couplings of phonons in the cross-plane direction. We have therefore investigated the presence of finite size effects by analyzing different systems with sizes ranging from $L=10a$ ($20$ monolayers, $N=4,000$) to $24a$ ($48$ monolayers, $N=55,296$). In Figs.~\ref{kmass}, \ref{kodis}, and \ref{klj}, we show multiple data points at some $w$-values, obtained for different system sizes. For the {\em S1} superlattice, we confirmed that the required number $P$ of repetitions becomes larger for smaller $w$~\cite{Landry_2008}: one period ($L=w$) only is adequate for $w \ge 20$, whereas four periods or more ($L \ge 4w$) are required for $w \le 8$. We have therefore employed four pattern repetitions ($L=4w$) for $10 \le w \le 12$ and two ($L=2w$) for $14 \le w \le 18$. This behaviour is simple to rationalize by inspecting the data in Fig.~\ref{kmass}, where the crossover between incoherent and coherent phonon transport occurs around $w^* \simeq 20$. In the coherent regime $w <20$, the wave character of the phonons becomes important, and therefore a larger number of repetitions is necessary to produce the coherent wave interference processes correctly. In contrast, smaller values of $P$ are needed (even $P=1$) in the incoherent regime $w>20$, where the incoherent particle nature of the phonons appears.

For the {\em S2} and {\em S3} superlattices the system size effects issue is much less pronounced than in the {\em S1} case. We can understand this behaviour by noticing that phonon tranport is mainly determined by the scattering processes in the disordered alloy layer in {\em S2}, and the blocking at the weak interface for {\em S3}. In both cases the missing long wavelength phonons, with $\lambda >L$, play very little role in phonon transport and finite system size effects are consequently negligible. We therefore used $P=1$ ($L = w$) for $w \ge 20$ and one or more repetitions ($L \ge w$) for $w < 20$, for both {\em S2} and {\em S3}.
\\

\noindent
{\bf Normal modes analysis.} We have characterized the superlattice vibrational states (superlattice phonons) by performing a standard normal-mode analysis~\cite{Ashcroft} with ARPACK~\cite{arpack}. We have diagonalized the dynamical (Hessian) matrix calculated at local minima of the potential energy landscape, and obtained eigenvalues $\lambda^k$ and eigenvectors (polarization vectors) $\mathbf{e}^k=\left\{\mathbf{e}^k_1,\ldots,\mathbf{e}^k_j,\ldots,\mathbf{e}^k_N\right\}$. Here, $j$ is the particle index, and $k=1,2,\ldots,3N-3$ is the eigenmode number, where we have disregarded the three vanishing Goldstone modes. The eigenvectors are normalized such that $\mathbf{e}^k \cdot \mathbf{e}^l = \sum_{j=1}^{N} (\mathbf{e}^k_j \cdot \mathbf{e}^l_j) = \delta_{kl}$, where $\delta_{kl}$ is the Kronecker delta function. The eigenfrequencies are next calculated as $\omega^k=\sqrt{\lambda^k}$, and the associated probability distribution (normalized histogram) directly provides the vDOS:
\begin{equation}
g(\omega) = \frac{1}{3N-3} \sum_{k=1}^{3N-3} \delta (\omega-\omega^k).
\label{vdos}
\end{equation}

In addition, from the eigenvector $\mathbf{e}^k$ we have defined the vibrational amplitudes of mode $k$ for layers $A$ and $B$:
\begin{equation}
E_{A(B)}^k = \sum_{j \in \text{layer}A(B)} \left( \mathbf{e}_j^k \cdot \mathbf{e}_j^k \right).
\label{amp}
\end{equation}
Note that $E^A_k+E^B_k =\mathbf{e}^k \cdot \mathbf{e}^k= 1$ for each $k$ and, therefore, $0 \le E_A^k, E_B^k \le 1$. Based on the values of $E_A^k$ and $E_B^k$, one can determine in which layer particles are more displaced (excited) according to the associated eigenvector $\mathbf{e}^k$. In particular, if $E^k_A \ge 0.5, E^k_B < 0.5$ ($E^k_A < 0.5, E^k_B \ge 0.5$), particles in layer $A$ ($B$) contribute more to mode $k$ than those in layer $B$ ($A$). In the case $E^k_A = E^k_B= 0.5$, particles in both layers contribute equivalently, and in a correlated manner. Note that the normal mode analysis provides us with the system vibrational states in the harmonic limit $T \rightarrow 0$ which, we believe, is an appropriate approximation for our case $T=10^{-2}$, where anharmonicities are weak. 
\\
\begin{acknowledgments}
We thank P. Keblinski for helpful correspondence. This work was supported by the Nanosciences Foundation of Grenoble. J.-L.~B is supported by the Institut Universitaire de France. Most of the computations presented in this work were performed using the Froggy platform of the CIMENT infrastructure (https://ciment.ujf-grenoble.fr), which is supported by the Rh\^{o}ne-Alpes region (GRANT CPER07\_13 CIRA) and the Equip@Meso project (reference ANR-10-EQPX-29-01) of the programme Investissements d'Avenir supervised by the Agence Nationale pour la Recherche.
\end{acknowledgments}

\begin{thebibliography}{66}
%
\makeatletter
\providecommand \@ifxundefined [1]{%
 \@ifx{#1\undefined}
}%
\providecommand \@ifnum [1]{%
 \ifnum #1\expandafter \@firstoftwo
 \else \expandafter \@secondoftwo
 \fi
}%
\providecommand \@ifx [1]{%
 \ifx #1\expandafter \@firstoftwo
 \else \expandafter \@secondoftwo
 \fi
}%
\providecommand \natexlab [1]{#1}%
\providecommand \enquote  [1]{``#1''}%
\providecommand \bibnamefont  [1]{#1}%
\providecommand \bibfnamefont [1]{#1}%
\providecommand \citenamefont [1]{#1}%
\providecommand \href@noop [0]{\@secondoftwo}%
\providecommand \href [0]{\begingroup \@sanitize@url \@href}%
\providecommand \@href[1]{\@@startlink{#1}\@@href}%
\providecommand \@@href[1]{\endgroup#1\@@endlink}%
\providecommand \@sanitize@url [0]{\catcode `\\12\catcode `\$12\catcode
  `\&12\catcode `\#12\catcode `\^12\catcode `\_12\catcode `\%12\relax}%
\providecommand \@@startlink[1]{}%
\providecommand \@@endlink[0]{}%
\providecommand \url  [0]{\begingroup\@sanitize@url \@url }%
\providecommand \@url [1]{\endgroup\@href {#1}{\urlprefix }}%
\providecommand \urlprefix  [0]{URL }%
\providecommand \Eprint [0]{\href }%
\providecommand \doibase [0]{http://dx.doi.org/}%
\providecommand \selectlanguage [0]{\@gobble}%
\providecommand \bibinfo  [0]{\@secondoftwo}%
\providecommand \bibfield  [0]{\@secondoftwo}%
\providecommand \translation [1]{[#1]}%
\providecommand \BibitemOpen [0]{}%
\providecommand \bibitemStop [0]{}%
\providecommand \bibitemNoStop [0]{.\EOS\space}%
\providecommand \EOS [0]{\spacefactor3000\relax}%
\providecommand \BibitemShut  [1]{\csname bibitem#1\endcsname}%
\let\auto@bib@innerbib\@empty
%
\bibitem [{\citenamefont {{\it et al.}}(2001)}]{Venkatasubramanian_2001}%
  \BibitemOpen
  \bibfield  {author} {\bibinfo {author} {\bibfnamefont
  {R.~Venkatasubramanian}\ \bibnamefont {{\it et al.}}},\ }\bibfield  {title}
  {\enquote {\bibinfo {title} {Thin-film thermoelectric devices with high
  room-temperature figures of merit},}\ }\href@noop {} {\bibfield  {journal}
  {\bibinfo  {journal} {Nature}\ }\textbf {\bibinfo {volume} {413}},\ \bibinfo
  {pages} {597--602} (\bibinfo {year} {2001})}\BibitemShut {NoStop}%
\bibitem [{\citenamefont {{\it et al.}}(2009{\natexlab{a}})}]{Minnich_2009}%
  \BibitemOpen
  \bibfield  {author} {\bibinfo {author} {\bibfnamefont {A.~J.~Minnich}\
  \bibnamefont {{\it et al.}}},\ }\bibfield  {title} {\enquote {\bibinfo
  {title} {Bulk nanostructured thermoelectric materials: current research and
  future prospects},}\ }\href {\doibase 10.1039/B822664B} {\bibfield  {journal}
  {\bibinfo  {journal} {Energy Environ. Sci.}\ }\textbf {\bibinfo {volume}
  {2}},\ \bibinfo {pages} {466--479} (\bibinfo {year}
  {2009}{\natexlab{a}})}\BibitemShut {NoStop}%
\bibitem [{\citenamefont {Maldovan}(2013)}]{Maldovan_2013}%
  \BibitemOpen
  \bibfield  {author} {\bibinfo {author} {\bibfnamefont {M.}~\bibnamefont
  {Maldovan}},\ }\bibfield  {title} {\enquote {\bibinfo {title} {Sound and heat
  revolutions in phononics},}\ }\href@noop {} {\bibfield  {journal} {\bibinfo
  {journal} {Nature}\ }\textbf {\bibinfo {volume} {503}},\ \bibinfo {pages}
  {209--217} (\bibinfo {year} {2013})}\BibitemShut {NoStop}%
\bibitem [{\citenamefont {Goodson}(2007)}]{Goodson_2007}%
  \BibitemOpen
  \bibfield  {author} {\bibinfo {author} {\bibfnamefont {K.~E.}\ \bibnamefont
  {Goodson}},\ }\bibfield  {title} {\enquote {\bibinfo {title} {Ordering up the
  minimum thermal conductivity of solids},}\ }\href {\doibase
  10.1126/science.1138067} {\bibfield  {journal} {\bibinfo  {journal}
  {Science}\ }\textbf {\bibinfo {volume} {315}},\ \bibinfo {pages} {342--343}
  (\bibinfo {year} {2007})}\BibitemShut {NoStop}%
\bibitem [{\citenamefont {Cahill}\ and\ \citenamefont
  {Pohl}(1988)}]{Cahill_1988}%
  \BibitemOpen
  \bibfield  {author} {\bibinfo {author} {\bibfnamefont {D.~G.}\ \bibnamefont
  {Cahill}}\ and\ \bibinfo {author} {\bibfnamefont {R.~O.}\ \bibnamefont
  {Pohl}},\ }\bibfield  {title} {\enquote {\bibinfo {title} {Lattice vibrations
  and heat transport in crystals and glasses},}\ }\href {\doibase
  10.1146/annurev.pc.39.100188.000521} {\bibfield  {journal} {\bibinfo
  {journal} {Annual Review of Physical Chemistry}\ }\textbf {\bibinfo {volume}
  {39}},\ \bibinfo {pages} {93--121} (\bibinfo {year} {1988})}\BibitemShut
  {NoStop}%
\bibitem [{\citenamefont {{\it et al.}}(1992)}]{Cahill_1992}%
  \BibitemOpen
  \bibfield  {author} {\bibinfo {author} {\bibfnamefont {D.~G.~Cahill}\
  \bibnamefont {{\it et al.}}},\ }\bibfield  {title} {\enquote {\bibinfo
  {title} {Lower limit to the thermal conductivity of disordered crystals},}\
  }\href {\doibase 10.1103/PhysRevB.46.6131} {\bibfield  {journal} {\bibinfo
  {journal} {Phys. Rev. B}\ }\textbf {\bibinfo {volume} {46}},\ \bibinfo
  {pages} {6131--6140} (\bibinfo {year} {1992})}\BibitemShut {NoStop}%
\bibitem [{\citenamefont {Allen}\ and\ \citenamefont
  {Feldman}(1993)}]{Allen_1993}%
  \BibitemOpen
  \bibfield  {author} {\bibinfo {author} {\bibfnamefont {P.~B.}\ \bibnamefont
  {Allen}}\ and\ \bibinfo {author} {\bibfnamefont {J.~L.}\ \bibnamefont
  {Feldman}},\ }\bibfield  {title} {\enquote {\bibinfo {title} {Thermal
  conductivity of disordered harmonic solids},}\ }\href {\doibase
  10.1103/PhysRevB.48.12581} {\bibfield  {journal} {\bibinfo  {journal} {Phys.
  Rev. B}\ }\textbf {\bibinfo {volume} {48}},\ \bibinfo {pages} {12581--12588}
  (\bibinfo {year} {1993})}\BibitemShut {NoStop}%
\bibitem [{\citenamefont {{\it et al.}}(2013{\natexlab{a}})}]{Mizuno2_2013}%
  \BibitemOpen
  \bibfield  {author} {\bibinfo {author} {\bibfnamefont {H.~Mizuno}\
  \bibnamefont {{\it et al.}}},\ }\bibfield  {title} {\enquote {\bibinfo
  {title} {Elastic heterogeneity, vibrational states, and thermal conductivity
  across an amorphisation transition},}\ }\href
  {http://stacks.iop.org/0295-5075/104/i=5/a=56001} {\bibfield  {journal}
  {\bibinfo  {journal} {EPL}\ }\textbf {\bibinfo {volume} {104}},\ \bibinfo
  {pages} {56001} (\bibinfo {year} {2013}{\natexlab{a}})}\BibitemShut {NoStop}%
\bibitem [{\citenamefont {{\it et al.}}()}]{Mizuno2_2014}%
  \BibitemOpen
  \bibfield  {author} {\bibinfo {author} {\bibfnamefont {H.~Mizuno}\
  \bibnamefont {{\it et al.}}},\ }\href@noop {} {\bibinfo  {journal} {in
  preparation}\ }\BibitemShut {NoStop}%
\bibitem [{\citenamefont {Kittel}(1996)}]{kettel}%
  \BibitemOpen
\bibfield  {journal} {  }\bibfield  {author} {\bibinfo {author} {\bibfnamefont
  {C.}~\bibnamefont {Kittel}},\ }\href@noop {} {\emph {\bibinfo {title}
  {Introduction to Solid State Physics}}},\ \bibinfo {edition} {7th}\ ed.\
  (\bibinfo  {publisher} {John Wiley and Sons, New York},\ \bibinfo {year}
  {1996})\BibitemShut {NoStop}%
\bibitem [{\citenamefont {{\it et al.}}(2013{\natexlab{b}})}]{Mizuno_2013}%
  \BibitemOpen
  \bibfield  {author} {\bibinfo {author} {\bibfnamefont {H.~Mizuno}\
  \bibnamefont {{\it et al.}}},\ }\bibfield  {title} {\enquote {\bibinfo
  {title} {Measuring spatial distribution of the local elastic modulus in
  glasses},}\ }\href {\doibase 10.1103/PhysRevE.87.042306} {\bibfield
  {journal} {\bibinfo  {journal} {Phys. Rev. E}\ }\textbf {\bibinfo {volume}
  {87}},\ \bibinfo {pages} {042306} (\bibinfo {year}
  {2013}{\natexlab{b}})}\BibitemShut {NoStop}%
\bibitem [{\citenamefont {{\it et al.}}(2011{\natexlab{a}})}]{Hopkins_2011}%
  \BibitemOpen
  \bibfield  {author} {\bibinfo {author} {\bibfnamefont {P.~E.~Hopkins}\
  \bibnamefont {{\it et al.}}},\ }\bibfield  {title} {\enquote {\bibinfo
  {title} {Reduction in the thermal conductivity of single crystalline silicon
  by phononic crystal patterning},}\ }\href {\doibase 10.1021/nl102918q}
  {\bibfield  {journal} {\bibinfo  {journal} {Nano Letters}\ }\textbf {\bibinfo
  {volume} {11}},\ \bibinfo {pages} {107--112} (\bibinfo {year}
  {2011}{\natexlab{a}})}\BibitemShut {NoStop}%
\bibitem [{\citenamefont {{\it et al.}}(1997)}]{Lee_1997}%
  \BibitemOpen
  \bibfield  {author} {\bibinfo {author} {\bibfnamefont {S.-M.~Lee}\
  \bibnamefont {{\it et al.}}},\ }\bibfield  {title} {\enquote {\bibinfo
  {title} {Thermal conductivity of $\text{Si}$-$\text{Ge}$ superlattices},}\
  }\href {\doibase http://dx.doi.org/10.1063/1.118755} {\bibfield  {journal}
  {\bibinfo  {journal} {Applied Physics Letters}\ }\textbf {\bibinfo {volume}
  {70}},\ \bibinfo {pages} {2957--2959} (\bibinfo {year} {1997})}\BibitemShut
  {NoStop}%
\bibitem [{\citenamefont {{\it et al.}}(2000{\natexlab{a}})}]{Volz2_2000}%
  \BibitemOpen
  \bibfield  {author} {\bibinfo {author} {\bibfnamefont {S.~Volz}\ \bibnamefont
  {{\it et al.}}},\ }\bibfield  {title} {\enquote {\bibinfo {title}
  {Computation of thermal conductivity of $\text{Si}$/$\text{Ge}$ superlattices
  by molecular dynamics techniques},}\ }\href {\doibase
  http://dx.doi.org/10.1016/S0026-2692(00)00064-1} {\bibfield  {journal}
  {\bibinfo  {journal} {Microelectronics Journal}\ }\textbf {\bibinfo {volume}
  {31}},\ \bibinfo {pages} {815--819} (\bibinfo {year}
  {2000}{\natexlab{a}})}\BibitemShut {NoStop}%
\bibitem [{\citenamefont {{\it et al.}}(1999)}]{Capinski_1999}%
  \BibitemOpen
  \bibfield  {author} {\bibinfo {author} {\bibfnamefont {W.~S.~Capinski}\
  \bibnamefont {{\it et al.}}},\ }\bibfield  {title} {\enquote {\bibinfo
  {title} {Thermal-conductivity measurements of $\text{GaAs}$/$\text{AlAs}$
  superlattices using a picosecond optical pump-and-probe technique},}\ }\href
  {\doibase 10.1103/PhysRevB.59.8105} {\bibfield  {journal} {\bibinfo
  {journal} {Phys. Rev. B}\ }\textbf {\bibinfo {volume} {59}},\ \bibinfo
  {pages} {8105--8113} (\bibinfo {year} {1999})}\BibitemShut {NoStop}%
\bibitem [{\citenamefont {Daly}\ and\ \citenamefont {Maris}(2002)}]{Daly_2002}%
  \BibitemOpen
  \bibfield  {author} {\bibinfo {author} {\bibfnamefont {B.~C.}\ \bibnamefont
  {Daly}}\ and\ \bibinfo {author} {\bibfnamefont {H.~J.}\ \bibnamefont
  {Maris}},\ }\bibfield  {title} {\enquote {\bibinfo {title} {Calculation of
  the thermal conductivity of superlattices by molecular dynamics
  simulation},}\ }\href {\doibase
  http://dx.doi.org/10.1016/S0921-4526(02)00476-3} {\bibfield  {journal}
  {\bibinfo  {journal} {Physica B: Condensed Matter}\ }\textbf {\bibinfo
  {volume} {316--317}},\ \bibinfo {pages} {247--249} (\bibinfo {year}
  {2002})}\BibitemShut {NoStop}%
\bibitem [{\citenamefont {Venkatasubramanian}(2000)}]{Venkatasubramanian_2000}%
  \BibitemOpen
  \bibfield  {author} {\bibinfo {author} {\bibfnamefont {R.}~\bibnamefont
  {Venkatasubramanian}},\ }\bibfield  {title} {\enquote {\bibinfo {title}
  {Lattice thermal conductivity reduction and phonon localizationlike behavior
  in superlattice structures},}\ }\href {\doibase 10.1103/PhysRevB.61.3091}
  {\bibfield  {journal} {\bibinfo  {journal} {Phys. Rev. B}\ }\textbf {\bibinfo
  {volume} {61}},\ \bibinfo {pages} {3091--3097} (\bibinfo {year}
  {2000})}\BibitemShut {NoStop}%
\bibitem [{\citenamefont {{\it et al.}}(2002{\natexlab{a}})}]{Yang_2002}%
  \BibitemOpen
  \bibfield  {author} {\bibinfo {author} {\bibfnamefont {B.~Yang}\ \bibnamefont
  {{\it et al.}}},\ }\bibfield  {title} {\enquote {\bibinfo {title}
  {Measurements of anisotropic thermoelectric properties in superlattices},}\
  }\href {\doibase http://dx.doi.org/10.1063/1.1515876} {\bibfield  {journal}
  {\bibinfo  {journal} {Applied Physics Letters}\ }\textbf {\bibinfo {volume}
  {81}},\ \bibinfo {pages} {3588--3590} (\bibinfo {year}
  {2002}{\natexlab{a}})}\BibitemShut {NoStop}%
\bibitem [{\citenamefont {{\it et
  al.}}(2007{\natexlab{a}})}]{Mavrokefalos_2007}%
  \BibitemOpen
  \bibfield  {author} {\bibinfo {author} {\bibfnamefont {A.~Mavrokefalos}\
  \bibnamefont {{\it et al.}}},\ }\bibfield  {title} {\enquote {\bibinfo
  {title} {In-plane thermal conductivity of disordered layered $\text{WSe}_2$
  and $(\text{W})_x(\text{WSe}_2)_y$ superlattice films},}\ }\href {\doibase
  http://dx.doi.org/10.1063/1.2800888} {\bibfield  {journal} {\bibinfo
  {journal} {Applied Physics Letters}\ }\textbf {\bibinfo {volume} {91}},\
  \bibinfo {eid} {171912} (\bibinfo {year} {2007}{\natexlab{a}})}\BibitemShut
  {NoStop}%
\bibitem [{\citenamefont {{\it et al.}}(2004)}]{Costescu_2004}%
  \BibitemOpen
  \bibfield  {author} {\bibinfo {author} {\bibfnamefont {R.~M.~Costescu}\
  \bibnamefont {{\it et al.}}},\ }\bibfield  {title} {\enquote {\bibinfo
  {title} {Ultra-low thermal conductivity in $\text{W}$/$\text{Al}_2\text{O}_3$
  nanolaminates},}\ }\href {\doibase 10.1126/science.1093711} {\bibfield
  {journal} {\bibinfo  {journal} {Science}\ }\textbf {\bibinfo {volume}
  {303}},\ \bibinfo {pages} {989--990} (\bibinfo {year} {2004})}\BibitemShut
  {NoStop}%
\bibitem [{\citenamefont {{\it et al.}}(2007{\natexlab{b}})}]{Chiritescu_2007}%
  \BibitemOpen
  \bibfield  {author} {\bibinfo {author} {\bibfnamefont {C.~Chiritescu}\
  \bibnamefont {{\it et al.}}},\ }\bibfield  {title} {\enquote {\bibinfo
  {title} {Ultralow thermal conductivity in disordered, layered $\text{WSe}_2$
  crystals},}\ }\href {\doibase 10.1126/science.1136494} {\bibfield  {journal}
  {\bibinfo  {journal} {Science}\ }\textbf {\bibinfo {volume} {315}},\ \bibinfo
  {pages} {351--353} (\bibinfo {year} {2007}{\natexlab{b}})}\BibitemShut
  {NoStop}%
\bibitem [{\citenamefont {{\it et al.}}(2010)}]{Pernot_2010}%
  \BibitemOpen
  \bibfield  {author} {\bibinfo {author} {\bibfnamefont {G.~Pernot}\
  \bibnamefont {{\it et al.}}},\ }\bibfield  {title} {\enquote {\bibinfo
  {title} {Precise control of thermal conductivity at the nanoscale through
  individual phonon-scattering barriers},}\ }\href@noop {} {\bibfield
  {journal} {\bibinfo  {journal} {Nature Mater.}\ }\textbf {\bibinfo {volume}
  {9}},\ \bibinfo {pages} {491--495} (\bibinfo {year} {2010})}\BibitemShut
  {NoStop}%
\bibitem [{\citenamefont {Simkin}\ and\ \citenamefont
  {Mahan}(2000)}]{Simkin_2000}%
  \BibitemOpen
  \bibfield  {author} {\bibinfo {author} {\bibfnamefont {M.~V.}\ \bibnamefont
  {Simkin}}\ and\ \bibinfo {author} {\bibfnamefont {G.~D.}\ \bibnamefont
  {Mahan}},\ }\bibfield  {title} {\enquote {\bibinfo {title} {Minimum thermal
  conductivity of superlattices},}\ }\href {\doibase
  10.1103/PhysRevLett.84.927} {\bibfield  {journal} {\bibinfo  {journal} {Phys.
  Rev. Lett.}\ }\textbf {\bibinfo {volume} {84}},\ \bibinfo {pages} {927--930}
  (\bibinfo {year} {2000})}\BibitemShut {NoStop}%
\bibitem [{\citenamefont {Yang}\ and\ \citenamefont {Chen}(2003)}]{Yang_2003}%
  \BibitemOpen
  \bibfield  {author} {\bibinfo {author} {\bibfnamefont {B.}~\bibnamefont
  {Yang}}\ and\ \bibinfo {author} {\bibfnamefont {G.}~\bibnamefont {Chen}},\
  }\bibfield  {title} {\enquote {\bibinfo {title} {Partially coherent phonon
  heat conduction in superlattices},}\ }\href {\doibase
  10.1103/PhysRevB.67.195311} {\bibfield  {journal} {\bibinfo  {journal} {Phys.
  Rev. B}\ }\textbf {\bibinfo {volume} {67}},\ \bibinfo {pages} {195311}
  (\bibinfo {year} {2003})}\BibitemShut {NoStop}%
\bibitem [{\citenamefont {Garg}\ and\ \citenamefont {Chen}(2013)}]{Garg_2013}%
  \BibitemOpen
  \bibfield  {author} {\bibinfo {author} {\bibfnamefont {J.}~\bibnamefont
  {Garg}}\ and\ \bibinfo {author} {\bibfnamefont {G.}~\bibnamefont {Chen}},\
  }\bibfield  {title} {\enquote {\bibinfo {title} {Minimum thermal conductivity
  in superlattices: A first-principles formalism},}\ }\href {\doibase
  10.1103/PhysRevB.87.140302} {\bibfield  {journal} {\bibinfo  {journal} {Phys.
  Rev. B}\ }\textbf {\bibinfo {volume} {87}},\ \bibinfo {pages} {140302}
  (\bibinfo {year} {2013})}\BibitemShut {NoStop}%
\bibitem [{\citenamefont {{\it et al.}}(2005)}]{Chen_2005}%
  \BibitemOpen
  \bibfield  {author} {\bibinfo {author} {\bibfnamefont {Y.~Chen}\ \bibnamefont
  {{\it et al.}}},\ }\bibfield  {title} {\enquote {\bibinfo {title} {Minimum
  superlattice thermal conductivity from molecular dynamics},}\ }\href
  {\doibase 10.1103/PhysRevB.72.174302} {\bibfield  {journal} {\bibinfo
  {journal} {Phys. Rev. B}\ }\textbf {\bibinfo {volume} {72}},\ \bibinfo
  {pages} {174302} (\bibinfo {year} {2005})}\BibitemShut {NoStop}%
\bibitem [{\citenamefont {{\it et al.}}(2007{\natexlab{c}})}]{Kawamura_2007}%
  \BibitemOpen
  \bibfield  {author} {\bibinfo {author} {\bibfnamefont {T.~Kawamura}\
  \bibnamefont {{\it et al.}}},\ }\bibfield  {title} {\enquote {\bibinfo
  {title} {An investigation of thermal conductivity of nitride-semiconductor
  nanostructures by molecular dynamics simulation},}\ }\href {\doibase
  http://dx.doi.org/10.1016/j.jcrysgro.2006.10.025} {\bibfield  {journal}
  {\bibinfo  {journal} {Journal of Crystal Growth}\ }\textbf {\bibinfo {volume}
  {298}},\ \bibinfo {pages} {251--253} (\bibinfo {year}
  {2007}{\natexlab{c}})}\BibitemShut {NoStop}%
\bibitem [{\citenamefont {{\it et al.}}(2008{\natexlab{a}})}]{Yang_2008}%
  \BibitemOpen
  \bibfield  {author} {\bibinfo {author} {\bibfnamefont {N.~Yang}\ \bibnamefont
  {{\it et al.}}},\ }\bibfield  {title} {\enquote {\bibinfo {title} {Ultralow
  thermal conductivity of isotope-doped silicon nanowires},}\ }\href {\doibase
  10.1021/nl0725998} {\bibfield  {journal} {\bibinfo  {journal} {Nano Letters}\
  }\textbf {\bibinfo {volume} {8}},\ \bibinfo {pages} {276--280} (\bibinfo
  {year} {2008}{\natexlab{a}})}\BibitemShut {NoStop}%
\bibitem [{\citenamefont {{\it et
  al.}}(2014{\natexlab{a}})}]{Ravichandran_2014}%
  \BibitemOpen
  \bibfield  {author} {\bibinfo {author} {\bibfnamefont {J.~Ravichandran}\
  \bibnamefont {{\it et al.}}},\ }\bibfield  {title} {\enquote {\bibinfo
  {title} {Crossover from incoherent to coherent phonon scattering in epitaxial
  oxide superlattices},}\ }\href@noop {} {\bibfield  {journal} {\bibinfo
  {journal} {Nature Mater.}\ }\textbf {\bibinfo {volume} {13}},\ \bibinfo
  {pages} {168--172} (\bibinfo {year} {2014}{\natexlab{a}})}\BibitemShut
  {NoStop}%
\bibitem [{\citenamefont {Chen}\ and\ \citenamefont {Neagu}(1997)}]{Chen_1997}%
  \BibitemOpen
  \bibfield  {author} {\bibinfo {author} {\bibfnamefont {G.}~\bibnamefont
  {Chen}}\ and\ \bibinfo {author} {\bibfnamefont {M.}~\bibnamefont {Neagu}},\
  }\bibfield  {title} {\enquote {\bibinfo {title} {Thermal conductivity and
  heat transfer in superlattices},}\ }\href {\doibase
  http://dx.doi.org/10.1063/1.120126} {\bibfield  {journal} {\bibinfo
  {journal} {Applied Physics Letters}\ }\textbf {\bibinfo {volume} {71}},\
  \bibinfo {pages} {2761--2763} (\bibinfo {year} {1997})}\BibitemShut {NoStop}%
\bibitem [{\citenamefont {Chen}(1998)}]{Chen_1998}%
  \BibitemOpen
  \bibfield  {author} {\bibinfo {author} {\bibfnamefont {G.}~\bibnamefont
  {Chen}},\ }\bibfield  {title} {\enquote {\bibinfo {title} {Thermal
  conductivity and ballistic-phonon transport in the cross-plane direction of
  superlattices},}\ }\href {\doibase 10.1103/PhysRevB.57.14958} {\bibfield
  {journal} {\bibinfo  {journal} {Phys. Rev. B}\ }\textbf {\bibinfo {volume}
  {57}},\ \bibinfo {pages} {14958--14973} (\bibinfo {year} {1998})}\BibitemShut
  {NoStop}%
\bibitem [{\citenamefont {{\it et al.}}(2000{\natexlab{b}})}]{Kim2_2000}%
  \BibitemOpen
  \bibfield  {author} {\bibinfo {author} {\bibfnamefont {E.-K.~Kim}\
  \bibnamefont {{\it et al.}}},\ }\bibfield  {title} {\enquote {\bibinfo
  {title} {Thermal boundary resistance at $\text{Ge}_2 \text{Sb}_2 \text{Te}_5
  / \text{ZnS:} \text{SiO}_2$ interface},}\ }\href@noop {} {\bibfield
  {journal} {\bibinfo  {journal} {Applied Physics Letters}\ }\textbf {\bibinfo
  {volume} {76}},\ \bibinfo {pages} {3864--3866} (\bibinfo {year}
  {2000}{\natexlab{b}})}\BibitemShut {NoStop}%
\bibitem [{\citenamefont {{\it et al.}}(2012{\natexlab{a}})}]{Lampin_2012}%
  \BibitemOpen
  \bibfield  {author} {\bibinfo {author} {\bibfnamefont {E.~Lampin}\
  \bibnamefont {{\it et al.}}},\ }\bibfield  {title} {\enquote {\bibinfo
  {title} {Thermal boundary resistance at silicon-silica interfaces by
  molecular dynamics simulations},}\ }\href@noop {} {\bibfield  {journal}
  {\bibinfo  {journal} {Applied Physics Letters}\ }\textbf {\bibinfo {volume}
  {100}},\ \bibinfo {eid} {131906} (\bibinfo {year}
  {2012}{\natexlab{a}})}\BibitemShut {NoStop}%
\bibitem [{\citenamefont {Nan}\ and\ \citenamefont
  {Birringer}(1998)}]{Nan_1998}%
  \BibitemOpen
  \bibfield  {author} {\bibinfo {author} {\bibfnamefont {C.-W.}\ \bibnamefont
  {Nan}}\ and\ \bibinfo {author} {\bibfnamefont {R.}~\bibnamefont
  {Birringer}},\ }\bibfield  {title} {\enquote {\bibinfo {title} {Determining
  the kapitza resistance and the thermal conductivity of polycrystals: A simple
  model},}\ }\href {\doibase 10.1103/PhysRevB.57.8264} {\bibfield  {journal}
  {\bibinfo  {journal} {Phys. Rev. B}\ }\textbf {\bibinfo {volume} {57}},\
  \bibinfo {pages} {8264--8268} (\bibinfo {year} {1998})}\BibitemShut {NoStop}%
\bibitem [{\citenamefont {Barrat}\ and\ \citenamefont
  {Chiaruttini}(2003)}]{BARRAT_2003}%
  \BibitemOpen
  \bibfield  {author} {\bibinfo {author} {\bibfnamefont {J.-L.}\ \bibnamefont
  {Barrat}}\ and\ \bibinfo {author} {\bibfnamefont {F.}~\bibnamefont
  {Chiaruttini}},\ }\bibfield  {title} {\enquote {\bibinfo {title} {Kapitza
  resistance at the liquid-solid interface},}\ }\href@noop {} {\bibfield
  {journal} {\bibinfo  {journal} {Molecular Physics}\ }\textbf {\bibinfo
  {volume} {101}},\ \bibinfo {pages} {1605--1610} (\bibinfo {year}
  {2003})}\BibitemShut {NoStop}%
\bibitem [{\citenamefont {{\it et al.}}(1988)}]{Tamura_1988}%
  \BibitemOpen
  \bibfield  {author} {\bibinfo {author} {\bibfnamefont {S.~Tamura}\
  \bibnamefont {{\it et al.}}},\ }\bibfield  {title} {\enquote {\bibinfo
  {title} {Acoustic-phonon propagation in superlattices},}\ }\href {\doibase
  10.1103/PhysRevB.38.1427} {\bibfield  {journal} {\bibinfo  {journal} {Phys.
  Rev. B}\ }\textbf {\bibinfo {volume} {38}},\ \bibinfo {pages} {1427--1449}
  (\bibinfo {year} {1988})}\BibitemShut {NoStop}%
\bibitem [{\citenamefont {Mizuno}\ and\ \citenamefont
  {i.~Tamura}(1992)}]{Mizuno_1992}%
  \BibitemOpen
  \bibfield  {author} {\bibinfo {author} {\bibfnamefont {S.}~\bibnamefont
  {Mizuno}}\ and\ \bibinfo {author} {\bibfnamefont {S.}~\bibnamefont
  {i.~Tamura}},\ }\bibfield  {title} {\enquote {\bibinfo {title} {Theory of
  acoustic-phonon transmission in finite-size superlattice systems},}\ }\href
  {\doibase 10.1103/PhysRevB.45.734} {\bibfield  {journal} {\bibinfo  {journal}
  {Phys. Rev. B}\ }\textbf {\bibinfo {volume} {45}},\ \bibinfo {pages}
  {734--741} (\bibinfo {year} {1992})}\BibitemShut {NoStop}%
\bibitem [{\citenamefont {Ren}\ and\ \citenamefont {Dow}(1982)}]{Ren_1982}%
  \BibitemOpen
  \bibfield  {author} {\bibinfo {author} {\bibfnamefont {S.~Y.}\ \bibnamefont
  {Ren}}\ and\ \bibinfo {author} {\bibfnamefont {J.~D.}\ \bibnamefont {Dow}},\
  }\bibfield  {title} {\enquote {\bibinfo {title} {Thermal conductivity of
  superlattices},}\ }\href {\doibase 10.1103/PhysRevB.25.3750} {\bibfield
  {journal} {\bibinfo  {journal} {Phys. Rev. B}\ }\textbf {\bibinfo {volume}
  {25}},\ \bibinfo {pages} {3750--3755} (\bibinfo {year} {1982})}\BibitemShut
  {NoStop}%
\bibitem [{\citenamefont {{\it et al.}}(2002{\natexlab{b}})}]{Daly2_2002}%
  \BibitemOpen
  \bibfield  {author} {\bibinfo {author} {\bibfnamefont {B.~C.~Daly}\
  \bibnamefont {{\it et al.}}},\ }\bibfield  {title} {\enquote {\bibinfo
  {title} {Molecular dynamics calculation of the thermal conductivity of
  superlattices},}\ }\href {\doibase 10.1103/PhysRevB.66.024301} {\bibfield
  {journal} {\bibinfo  {journal} {Phys. Rev. B}\ }\textbf {\bibinfo {volume}
  {66}},\ \bibinfo {pages} {024301} (\bibinfo {year}
  {2002}{\natexlab{b}})}\BibitemShut {NoStop}%
\bibitem [{\citenamefont {{\it et al.}}(2003{\natexlab{a}})}]{Imamura_2003}%
  \BibitemOpen
  \bibfield  {author} {\bibinfo {author} {\bibfnamefont {K.~Imamura}\
  \bibnamefont {{\it et al.}}},\ }\bibfield  {title} {\enquote {\bibinfo
  {title} {Lattice thermal conductivity in superlattices: molecular dynamics
  calculations with a heat reservoir method},}\ }\href
  {http://stacks.iop.org/0953-8984/15/i=50/a=002} {\bibfield  {journal}
  {\bibinfo  {journal} {Journal of Physics: Condensed Matter}\ }\textbf
  {\bibinfo {volume} {15}},\ \bibinfo {pages} {8679--8690} (\bibinfo {year}
  {2003}{\natexlab{a}})}\BibitemShut {NoStop}%
\bibitem [{\citenamefont {{\it et al.}}(2003{\natexlab{b}})}]{Daly_2003}%
  \BibitemOpen
  \bibfield  {author} {\bibinfo {author} {\bibfnamefont {B.~C.~Daly}\
  \bibnamefont {{\it et al.}}},\ }\bibfield  {title} {\enquote {\bibinfo
  {title} {Molecular dynamics calculation of the in-plane thermal conductivity
  of $\text{GaAs}$/$\text{AlAs}$ superlattices},}\ }\href {\doibase
  10.1103/PhysRevB.67.033308} {\bibfield  {journal} {\bibinfo  {journal} {Phys.
  Rev. B}\ }\textbf {\bibinfo {volume} {67}},\ \bibinfo {pages} {033308}
  (\bibinfo {year} {2003}{\natexlab{b}})}\BibitemShut {NoStop}%
\bibitem [{\citenamefont {Landry}\ and\ \citenamefont
  {McGaughey}(2009)}]{Landry_2009}%
  \BibitemOpen
  \bibfield  {author} {\bibinfo {author} {\bibfnamefont {E.~S.}\ \bibnamefont
  {Landry}}\ and\ \bibinfo {author} {\bibfnamefont {A.~J.~H.}\ \bibnamefont
  {McGaughey}},\ }\bibfield  {title} {\enquote {\bibinfo {title} {Effect of
  interfacial species mixing on phonon transport in semiconductor
  superlattices},}\ }\href {\doibase 10.1103/PhysRevB.79.075316} {\bibfield
  {journal} {\bibinfo  {journal} {Phys. Rev. B}\ }\textbf {\bibinfo {volume}
  {79}},\ \bibinfo {pages} {075316} (\bibinfo {year} {2009})}\BibitemShut
  {NoStop}%
\bibitem [{\citenamefont {{\it et al.}}(2013{\natexlab{c}})}]{Huberman_2013}%
  \BibitemOpen
  \bibfield  {author} {\bibinfo {author} {\bibfnamefont {S.~C.~Huberman}\
  \bibnamefont {{\it et al.}}},\ }\bibfield  {title} {\enquote {\bibinfo
  {title} {Disruption of superlattice phonons by interfacial mixing},}\ }\href
  {\doibase 10.1103/PhysRevB.88.155311} {\bibfield  {journal} {\bibinfo
  {journal} {Phys. Rev. B}\ }\textbf {\bibinfo {volume} {88}},\ \bibinfo
  {pages} {155311} (\bibinfo {year} {2013}{\natexlab{c}})}\BibitemShut
  {NoStop}%
\bibitem [{\citenamefont {{\it et
  al.}}(2009{\natexlab{b}})}]{Termentzidis_2009}%
  \BibitemOpen
  \bibfield  {author} {\bibinfo {author} {\bibfnamefont {K.~Termentzidis}\
  \bibnamefont {{\it et al.}}},\ }\bibfield  {title} {\enquote {\bibinfo
  {title} {Nonequilibrium molecular dynamics simulation of the in-plane thermal
  conductivity of superlattices with rough interfaces},}\ }\href {\doibase
  10.1103/PhysRevB.79.214307} {\bibfield  {journal} {\bibinfo  {journal} {Phys.
  Rev. B}\ }\textbf {\bibinfo {volume} {79}},\ \bibinfo {pages} {214307}
  (\bibinfo {year} {2009}{\natexlab{b}})}\BibitemShut {NoStop}%
\bibitem [{\citenamefont {{\it et
  al.}}(2011{\natexlab{b}})}]{Termentzidis_2011}%
  \BibitemOpen
  \bibfield  {author} {\bibinfo {author} {\bibfnamefont {K.~Termentzidis}\
  \bibnamefont {{\it et al.}}},\ }\bibfield  {title} {\enquote {\bibinfo
  {title} {Cross-plane thermal conductivity of superlattices with rough
  interfaces using equilibrium and non-equilibrium molecular dynamics},}\
  }\href {\doibase http://dx.doi.org/10.1016/j.ijheatmasstransfer.2011.01.001}
  {\bibfield  {journal} {\bibinfo  {journal} {International Journal of Heat and
  Mass Transfer}\ }\textbf {\bibinfo {volume} {54}},\ \bibinfo {pages}
  {2014--2020} (\bibinfo {year} {2011}{\natexlab{b}})}\BibitemShut {NoStop}%
\bibitem [{\citenamefont {{\it et al.}}(2011{\natexlab{c}})}]{Hsieh_2011}%
  \BibitemOpen
  \bibfield  {author} {\bibinfo {author} {\bibfnamefont {W.-P.~Hsieh}\
  \bibnamefont {{\it et al.}}},\ }\bibfield  {title} {\enquote {\bibinfo
  {title} {Pressure tuning of the thermal conductance of weak interfaces},}\
  }\href {\doibase 10.1103/PhysRevB.84.184107} {\bibfield  {journal} {\bibinfo
  {journal} {Phys. Rev. B}\ }\textbf {\bibinfo {volume} {84}},\ \bibinfo
  {pages} {184107} (\bibinfo {year} {2011}{\natexlab{c}})}\BibitemShut
  {NoStop}%
\bibitem [{\citenamefont {{\it et al.}}(2011{\natexlab{d}})}]{Shen_2011}%
  \BibitemOpen
  \bibfield  {author} {\bibinfo {author} {\bibfnamefont {M.~Shen}\ \bibnamefont
  {{\it et al.}}},\ }\bibfield  {title} {\enquote {\bibinfo {title} {Bonding
  and pressure-tunable interfacial thermal conductance},}\ }\href {\doibase
  10.1103/PhysRevB.84.195432} {\bibfield  {journal} {\bibinfo  {journal} {Phys.
  Rev. B}\ }\textbf {\bibinfo {volume} {84}},\ \bibinfo {pages} {195432}
  (\bibinfo {year} {2011}{\natexlab{d}})}\BibitemShut {NoStop}%
\bibitem [{\citenamefont {{\it et al.}}(2012{\natexlab{b}})}]{Losego_2012}%
  \BibitemOpen
  \bibfield  {author} {\bibinfo {author} {\bibfnamefont {M.~D.~Losego}\
  \bibnamefont {{\it et al.}}},\ }\bibfield  {title} {\enquote {\bibinfo
  {title} {Effects of chemical bonding on heat transport across interfaces},}\
  }\href@noop {} {\bibfield  {journal} {\bibinfo  {journal} {Nature Mater.}\
  }\textbf {\bibinfo {volume} {11}},\ \bibinfo {pages} {502--506} (\bibinfo
  {year} {2012}{\natexlab{b}})}\BibitemShut {NoStop}%
\bibitem [{\citenamefont {{\it et al.}}(2013{\natexlab{d}})}]{Wei_2013}%
  \BibitemOpen
  \bibfield  {author} {\bibinfo {author} {\bibfnamefont {Z.~Wei}\ \bibnamefont
  {{\it et al.}}},\ }\bibfield  {title} {\enquote {\bibinfo {title} {Negative
  correlation between in-plane bonding strength and cross-plane thermal
  conductivity in a model layered material},}\ }\href {\doibase
  http://dx.doi.org/10.1063/1.4773372} {\bibfield  {journal} {\bibinfo
  {journal} {Applied Physics Letters}\ }\textbf {\bibinfo {volume} {102}},\
  \bibinfo {eid} {011901} (\bibinfo {year} {2013}{\natexlab{d}})}\BibitemShut
  {NoStop}%
\bibitem [{\citenamefont {McGaughey}\ and\ \citenamefont
  {Kaviany}(2006)}]{McGaughey}%
  \BibitemOpen
  \bibfield  {author} {\bibinfo {author} {\bibfnamefont {A.~J.~H.}\
  \bibnamefont {McGaughey}}\ and\ \bibinfo {author} {\bibfnamefont
  {M.}~\bibnamefont {Kaviany}},\ }\href@noop {} {\emph {\bibinfo {title}
  {Advances in Heat Transfer, edited by G. Greene, Y. Cho, J. Hartnett, and A.
  Bar-Cohen}}},\ Vol.~\bibinfo {volume} {39}\ (\bibinfo  {publisher} {Elsevier,
  New York},\ \bibinfo {year} {2006})\ pp.\ \bibinfo {pages}
  {169--255}\BibitemShut {NoStop}%
\bibitem [{\citenamefont {{\it et al.}}(2008{\natexlab{b}})}]{Landry_2008}%
  \BibitemOpen
  \bibfield  {author} {\bibinfo {author} {\bibfnamefont {E.~S.~Landry}\
  \bibnamefont {{\it et al.}}},\ }\bibfield  {title} {\enquote {\bibinfo
  {title} {Complex superlattice unit cell designs for reduced thermal
  conductivity},}\ }\href {\doibase 10.1103/PhysRevB.77.184302} {\bibfield
  {journal} {\bibinfo  {journal} {Phys. Rev. B}\ }\textbf {\bibinfo {volume}
  {77}},\ \bibinfo {pages} {184302} (\bibinfo {year}
  {2008}{\natexlab{b}})}\BibitemShut {NoStop}%
\bibitem [{\citenamefont {Ashcroft}\ and\ \citenamefont
  {Mermin}(1976)}]{Ashcroft}%
  \BibitemOpen
  \bibfield  {author} {\bibinfo {author} {\bibfnamefont {N.~W.}\ \bibnamefont
  {Ashcroft}}\ and\ \bibinfo {author} {\bibfnamefont {N.~D.}\ \bibnamefont
  {Mermin}},\ }\href@noop {} {\emph {\bibinfo {title} {Solid State Physics}}}\
  (\bibinfo  {publisher} {Harcourt College Publishers, New York},\ \bibinfo
  {year} {1976})\BibitemShut {NoStop}%
\bibitem [{\citenamefont {{\it et al.}}(2012{\natexlab{c}})}]{Yang_2012}%
  \BibitemOpen
  \bibfield  {author} {\bibinfo {author} {\bibfnamefont {L.~Yang}\ \bibnamefont
  {{\it et al.}}},\ }\bibfield  {title} {\enquote {\bibinfo {title} {Reduction
  of thermal conductivity by nanoscale 3d phononic crystal},}\ }\href@noop {}
  {\bibfield  {journal} {\bibinfo  {journal} {Nature Scientific Reports}\
  }\textbf {\bibinfo {volume} {3}},\ \bibinfo {pages} {1143} (\bibinfo {year}
  {2012}{\natexlab{c}})}\BibitemShut {NoStop}%
\bibitem [{\citenamefont {{\it et al.}}(2003{\natexlab{c}})}]{Li_2003}%
  \BibitemOpen
  \bibfield  {author} {\bibinfo {author} {\bibfnamefont {D.~Li}\ \bibnamefont
  {{\it et al.}}},\ }\bibfield  {title} {\enquote {\bibinfo {title} {Thermal
  conductivity of $\text{Si}$/$\text{SiGe}$ superlattice nanowires},}\ }\href
  {\doibase http://dx.doi.org/10.1063/1.1619221} {\bibfield  {journal}
  {\bibinfo  {journal} {Applied Physics Letters}\ }\textbf {\bibinfo {volume}
  {83}},\ \bibinfo {pages} {3186--3188} (\bibinfo {year}
  {2003}{\natexlab{c}})}\BibitemShut {NoStop}%
\bibitem [{\citenamefont {Monaco}\ and\ \citenamefont
  {Mossa}(2009)}]{monaco2_2009}%
  \BibitemOpen
  \bibfield  {author} {\bibinfo {author} {\bibfnamefont {G.}~\bibnamefont
  {Monaco}}\ and\ \bibinfo {author} {\bibfnamefont {S.}~\bibnamefont {Mossa}},\
  }\bibfield  {title} {\enquote {\bibinfo {title} {Anomalous properties of the
  acoustic excitations in glasses on the mesoscopic length scale},}\
  }\href@noop {} {\bibfield  {journal} {\bibinfo  {journal} {Proc. Natl. Acad.
  Sci. USA}\ }\textbf {\bibinfo {volume} {106}},\ \bibinfo {pages}
  {16907--16912} (\bibinfo {year} {2009})}\BibitemShut {NoStop}%
\bibitem [{\citenamefont {{\it et al.}}(2014{\natexlab{b}})}]{Mizuno_2014}%
  \BibitemOpen
  \bibfield  {author} {\bibinfo {author} {\bibfnamefont {H.~Mizuno}\
  \bibnamefont {{\it et al.}}},\ }\bibfield  {title} {\enquote {\bibinfo
  {title} {Acoustic excitations and elastic heterogeneities in disordered
  solids},}\ }\href
  {http://www.pnas.org/content/early/2014/07/31/1409490111.abstract} {\bibfield
   {journal} {\bibinfo  {journal} {Proc. Natl. Acad. Sci. USA}\ }\textbf
  {\bibinfo {volume} {111}},\ \bibinfo {pages} {11949--11954} (\bibinfo {year}
  {2014}{\natexlab{b}})}\BibitemShut {NoStop}%
\bibitem [{\citenamefont {{\it et al.}}(2012{\natexlab{d}})}]{Luckyanova_2012}%
  \BibitemOpen
  \bibfield  {author} {\bibinfo {author} {\bibfnamefont {M.~N.~Luckyanova}\
  \bibnamefont {{\it et al.}}},\ }\bibfield  {title} {\enquote {\bibinfo
  {title} {Coherent phonon heat conduction in superlattices},}\ }\href
  {\doibase 10.1126/science.1225549} {\bibfield  {journal} {\bibinfo  {journal}
  {Science}\ }\textbf {\bibinfo {volume} {338}},\ \bibinfo {pages} {936--939}
  (\bibinfo {year} {2012}{\natexlab{d}})}\BibitemShut {NoStop}%
\bibitem [{\citenamefont {{\it et al.}}(2006)}]{Woochul_2006}%
  \BibitemOpen
  \bibfield  {author} {\bibinfo {author} {\bibfnamefont {W.~Kim}\ \bibnamefont
  {{\it et al.}}},\ }\href {\doibase 10.1103/PhysRevLett.96.045901} {\bibfield
  {journal} {\bibinfo  {journal} {Phys. Rev. Lett.}\ }\textbf {\bibinfo
  {volume} {96}},\ \bibinfo {pages} {045901} (\bibinfo {year}
  {2006})}\BibitemShut {NoStop}%
\bibitem [{\citenamefont {Zhang}\ and\ \citenamefont
  {Minnich}(2014)}]{Zhang_2014}%
  \BibitemOpen
  \bibfield  {author} {\bibinfo {author} {\bibfnamefont {H.}~\bibnamefont
  {Zhang}}\ and\ \bibinfo {author} {\bibfnamefont {A.~J.}\ \bibnamefont
  {Minnich}},\ }\bibfield  {title} {\enquote {\bibinfo {title} {The best
  nanoparticle size distribution for minimum thermal conductivity},}\
  }\href@noop {} {\bibfield  {journal} {\bibinfo  {journal} {arXiv:1404.1438}\
  } (\bibinfo {year} {2014})}\BibitemShut {NoStop}%
\bibitem [{\citenamefont {{\it et al.}}(2011{\natexlab{e}})}]{Nika_2011}%
  \BibitemOpen
  \bibfield  {author} {\bibinfo {author} {\bibfnamefont {D.~L.~Nika}\
  \bibnamefont {{\it et al.}}},\ }\bibfield  {title} {\enquote {\bibinfo
  {title} {Reduction of lattice thermal conductivity in one-dimensional
  quantum-dot superlattices due to phonon filtering},}\ }\href {\doibase
  10.1103/PhysRevB.84.165415} {\bibfield  {journal} {\bibinfo  {journal} {Phys.
  Rev. B}\ }\textbf {\bibinfo {volume} {84}},\ \bibinfo {pages} {165415}
  (\bibinfo {year} {2011}{\natexlab{e}})}\BibitemShut {NoStop}%
\bibitem [{\citenamefont {{\it et al.}}(2014{\natexlab{c}})}]{Yang_2014}%
  \BibitemOpen
  \bibfield  {author} {\bibinfo {author} {\bibfnamefont {L.~Yang}\ \bibnamefont
  {{\it et al.}}},\ }\bibfield  {title} {\enquote {\bibinfo {title} {Extreme
  low thermal conductivity in nanoscale 3$\text{D}$ $\text{Si}$ phononic
  crystal with spherical pores},}\ }\href {\doibase 10.1021/nl403750s}
  {\bibfield  {journal} {\bibinfo  {journal} {Nano Letters}\ }\textbf {\bibinfo
  {volume} {14}},\ \bibinfo {pages} {1734--1738} (\bibinfo {year}
  {2014}{\natexlab{c}})}\BibitemShut {NoStop}%
\bibitem [{\citenamefont {Plimpton}(1995)}]{Plimpton_1995}%
  \BibitemOpen
  \bibfield  {author} {\bibinfo {author} {\bibfnamefont {S.}~\bibnamefont
  {Plimpton}},\ }\bibfield  {title} {\enquote {\bibinfo {title} {Fast parallel
  algorithms for short-range molecular dynamics},}\ }\href {\doibase
  http://dx.doi.org/10.1006/jcph.1995.1039} {\bibfield  {journal} {\bibinfo
  {journal} {Journal of Computational Physics}\ }\textbf {\bibinfo {volume}
  {117}},\ \bibinfo {pages} {1--19} (\bibinfo {year} {1995})}\BibitemShut
  {NoStop}%
\bibitem [{lam()}]{lammps}%
  \BibitemOpen
  \href@noop {} {}\bibinfo {note} {Http://lammps.sandia.gov.}\BibitemShut
  {Stop}%
\bibitem [{\citenamefont {{\it et al.}}(2009{\natexlab{c}})}]{Turney_2009}%
  \BibitemOpen
  \bibfield  {author} {\bibinfo {author} {\bibfnamefont {J.~E.~Turney}\
  \bibnamefont {{\it et al.}}},\ }\bibfield  {title} {\enquote {\bibinfo
  {title} {Assessing the applicability of quantum corrections to classical
  thermal conductivity predictions},}\ }\href {\doibase
  10.1103/PhysRevB.79.224305} {\bibfield  {journal} {\bibinfo  {journal} {Phys.
  Rev. B}\ }\textbf {\bibinfo {volume} {79}},\ \bibinfo {pages} {224305}
  (\bibinfo {year} {2009}{\natexlab{c}})}\BibitemShut {NoStop}%
\bibitem [{\citenamefont {{\it et al.}}(2014{\natexlab{d}})}]{Bedoya_2014}%
  \BibitemOpen
  \bibfield  {author} {\bibinfo {author} {\bibfnamefont {O.~N.
  Bedoya-Martinez}\ \bibnamefont {{\it et al.}}},\ }\bibfield  {title}
  {\enquote {\bibinfo {title} {Computation of the thermal conductivity using
  methods based on classical and quantum molecular dynamics},}\ }\href
  {\doibase 10.1103/PhysRevB.89.014303} {\bibfield  {journal} {\bibinfo
  {journal} {Phys. Rev. B}\ }\textbf {\bibinfo {volume} {89}},\ \bibinfo
  {pages} {014303} (\bibinfo {year} {2014}{\natexlab{d}})}\BibitemShut
  {NoStop}%
\bibitem [{arp()}]{arpack}%
  \BibitemOpen
  \href@noop {} {}\bibinfo {note}
  {Http://www.caam.rice.edu/software/ARPACK.}\BibitemShut {Stop}%
\end{thebibliography}
\end{document}